\documentclass[12pt,preprint]{aastex}
\usepackage{epsfig}
\usepackage{lscape}
\newcommand{\ergsscm}{ergs~s$^{-1}$~cm$^{-2}$}
\newcommand{\HST}{{\sl HST}}
\newcommand{\CXO}{{\sl CXO}}

\slugcomment{Accepted by the {\it The Astrophysical Journal}}

\shorttitle{\HST\ Morphology of \CXO\ Deep Field Sources}
\shortauthors{Grogin et al.}

\received{2003 April 11}
\begin{document}
\title{{\sl Hubble Space Telescope} Imaging in the {\sl Chandra} Deep Field South: III.
Quantitative Morphology of the 1~Million Second {\sl Chandra} Counterparts and Comparison
with the Field Population
\footnote{Based on observations made with the NASA/ESA Hubble Space Telescope,
which is operated by the Association of Universities for Research in Astronomy,
Inc., under NASA contract \hbox{NAS 5-26555}.}}

\author{N.~A.~Grogin\altaffilmark{1}, A.~M.~Koekemoer\altaffilmark{2},
E.~J.~Schreier\altaffilmark{2,3}, J.~Bergeron\altaffilmark{4},
R.~Giacconi\altaffilmark{1,3},
G.~Hasinger\altaffilmark{5},
L.~Kewley\altaffilmark{6}, C.~Norman\altaffilmark{1},
P.~Rosati\altaffilmark{4}, P.~Tozzi\altaffilmark{7}, and
A.~Zirm\altaffilmark{1}} 
\altaffiltext{1}{Department of Physics and
Astronomy, Johns Hopkins University, Baltimore, MD 21218, USA}
\altaffiltext{2}{Space Telescope Science Institute, 3700 San Martin
Drive, Baltimore, MD 21218, USA} 
\altaffiltext{3}{Associated
Universities, Inc., 1400 16th Street NW, Washington, DC 20036, USA}
\altaffiltext{4}{European Southern Observatory,
Karl-Schwarzschild-Strasse 2, Garching, D-85748, Germany}
\altaffiltext{5}{Max-Planck-Institut F\"ur extraterrestrische Physik,
Giessenbachstrasse PF 1312, Garching D-85748, Germany}
\altaffiltext{6}{Harvard-Smithsonian Center for Astrophysics, 60
Garden St., Cambridge, MA 02138,USA} 
\altaffiltext{7}{Osservatorio
Astronomico di Trieste, Via G. B. Tiepolo 11, 34131 Trieste, Italy}

\begin{abstract}
We present quantitative morphological analyses of 37 \HST/WFPC2
counterparts of X-ray sources in the 1 Ms Chandra Deep Field--South
(CDFS).  We investigate: 1) one-dimensional surface brightness profiles via
isophotal ellipse fitting; 2) two-dimensional, PSF-convolved, bulge+disk+nucleus
profile-fitting; 3) asymmetry and concentration indices compared with
all $\sim\!3000$ sources in our three WFPC2 fields; and 4)
near-neighbor analyses comparing local environments of X-ray sources
versus the field control sample.  Significant nuclear point-source
optical components appear in roughly half of the resolved \HST/WFPC2
counterparts, showing a narrow range of $F_X/F_{opt,nuc}$ consistent
with the several \HST-unresolved X-ray sources (putative type~1 active
galactic nuclei [AGNs])
in our fields.  We infer roughly half of the \HST/WFPC2 counterparts
host unobscured AGN, comparable to analogous low-redshift AGN samples
and suggesting no steep decline in the type~1/type~2 ratio out to the
redshifts $z\sim 0.5-1$ typical of our sources.  The concentration
indices of the CDFS counterparts are clearly larger on average than
those of the field distribution, at 5-sigma, suggesting that the
strong correlation between central black hole mass and host galaxy
properties (including concentration index) observed in nearby galaxies
is already evident by $z\sim 0.5-1$.  By contrast, the asymmetry index
distribution of the 21 resolved CDFS sources at $I<23$ is
indistinguishable from the $I<23$ field.  Moreover, the frequency of
$I<23$ near neighbors around the CDFS counterparts is not
significantly different from the field sample.  These results,
combined with previous similar findings for local samples, suggest
that recent merger/interaction history is not a good indicator of AGN
activity over a substantial range of look-back time.
\end{abstract}

\keywords{galaxies: structure ---
        X-rays: galaxies ---
        galaxies: active ---
        galaxies: fundamental parameters ---
        surveys}

\section{Introduction}
It is becoming clear that a proper understanding of galaxy formation
and evolution must address the role played by the supermassive black
holes (SBH) now suspected to be present in most galaxy nuclei (e.g.,
Magorrian et al.~1998) and responsible for the AGN phenomenon.  In the
local universe, the importance of the SBH--host-galaxy relationship
has been highlighted recently by the tight correlation of SBH mass
with the host bulge velocity dispersion (Gebhardt et al.~2000;
Ferrarese \& Merritt 2000) and concentration index (Graham et
al.~2001).  It is of great interest to probe the AGN--host-galaxy
relationship out to higher redshifts, where, for example, the strong evolution
observed at the low end of the AGN luminosity function by {\sl ROSAT}
deep X-ray surveys (Hasinger et al.~1999; Miyaji et al.~2000) has been
associated with an increased rate of galaxy interactions and
distorted/irregular morphologies in the early universe (Lilly et
al.~1998; Abraham et al.~1999).  

The recent {\sl Chandra X-ray Observatory} (\CXO) Deepest Field
surveys have extended the {\sl ROSAT} results to fainter fluxes and
harder X-ray energies (Cowie et al.~2003; Giacconi et al.~2002; Brandt
et al.~2001; Hornschemeier et al.~2001; Mushotzky et al.~2000; Tozzi
et al.~2001).  These studies have also confirmed that the cosmic X-ray
background is largely attributable to moderate-luminosity AGN ($L_X
\sim 10^{43-44}$ erg s${}^{-1}$) at moderate to high redshifts
($z\gtrsim0.5$), many with significant obscuration by the host galaxy.
This underscores the desirability of studying the AGN--host connection
with samples detected at hard X-ray energies, where there is much less
bias against the dust obscuration that segregates AGN into type~1
(unobscured) and type~2 (obscured) in the standard unified model
(e.g., Urry \& Padovani 1995).  Although much has been learned from
studies of AGN and their hosts at low redshift, and from comparisons
of the evolution of higher-luminosity AGN and field galaxies at high
redshifts, the fundamental question of whether nuclear activity drives
galaxy evolution, or vice versa, can only be fully addressed by
obtaining high-resolution multi-wavelength observations of {\em
typical} AGN {\em and} their hosts over a {\em wide range} of
redshifts.  The combination of {\sl Hubble Space Telescope} (\HST) and
\CXO\ deep imaging offers a unique opportunity to satisfy these
requirements: the {\sl Chandra} Deepest Fields contain the requisite
large, unbiased sample of distant low- to moderate-luminosity AGN with
the high positional accuracy ($\lesssim0\farcs5$) necessary for
unambiguous association with their faint host galaxies; subsequent
high-resolution \HST\ images of these regions then permit the detailed
investigation of their host galaxy morphologies and environments.

To this end, we obtained the first \HST\ imaging within the Chandra
Deep Field South (hereafter CDFS; Giacconi et al.~2002): three
moderately-deep exposures using the Wide Field Planetary Camera~2
(WFPC2) in $V$ and $I$.  In Paper I of this series (Schreier et
al.~2001), we identified unambiguous {\sl HST} counterparts for
$>90$\% of the 26 CDFS sources in the initial 300~ksec exposure
(Tozzi et al.~2001), and discussed the heterogeneous nature of this
population as indicated by the joint X-ray and optical photometry.
Paper II (Koekemoer et al.~2002) extended this analysis to the larger
population of 40 CDFS sources revealed at the 1~Ms \CXO\ depth,
which is still current at the time of this writing.  Again we found a
$>90$\% fraction with unambiguous \HST\ counterparts, and the combined
X-ray and optical properties of this expanded sample largely
reinforced the conclusions of Paper I: the optical counterparts to the
X-ray sources are divided into two distinct populations, namely 1)~an
optically faint group with relatively blue colors, similar to the
faint blue field galaxy population, and 2)~an optically brighter
group, including resolved galaxies with average colors significantly
redder than the corresponding bright field galaxy population.  The
brighter objects comprise a wide range of types, including early and
late type galaxies, starbursts, and AGN, most at moderate redshifts
($z\sim 0.4$--1.0; Szokoly et al., in prep.).  The faint blue X-ray
population is consistent with expected numbers and characteristics of
low- to moderate-luminosity type~2 AGN at the quasar epoch ($z \sim
2$--3).
 
In the present paper, we undertake a detailed morphological analysis
of the CDFS optical counterparts introduced in Papers I and II.  In
\S\ref{obsrcsec} we review the \HST\ and \CXO\ imaging of the CDFS,
introduce the full \HST/WFPC2 source list, and review the matching of
sources common to both sets of observations.  We then present a
variety of morphological analyses of the \HST\ data in
\S\ref{morphansec}, including isophotal ellipse fitting and
point-spread function (PSF) convolved two-dimensional surface-brightness modeling
of the \CXO\ counterparts, as well as determinations of asymmetry and
concentration indices for the entire \HST\ source list in order to
contrast the \CXO\ counterparts with the much more numerous field
population in our WFPC2 frames.  We discuss the implications of these
findings in \S\ref{discsec} and conclude in \S\ref{concsec}.
Throughout this paper we adopt a cosmology with $\Omega_{\rm M} = 0.3$, 
$\Omega_\Lambda = 0.7$, and $H_0 = 70$~km~s$^{-1}$~Mpc$^{-1}$.

\section{Observations and Source Catalogs \label{obsrcsec}}
In \S\ref{hstobs} and \S\ref{cxoobs} we provide a brief summary of the
\HST\ and \CXO\ observations used in this study.  We refer the reader
to Paper II and to Giacconi et al.~(2002), respectively, for detailed
descriptions of the \HST\ and \CXO\ observations and reduction.  In
\S\ref{srclist} we present the combined source catalog from our three
\HST/WFPC2 fields and summarize the optical--X-ray source matching
procedure detailed in Paper II.  We conclude this section with the
catalog of 37 optical--X-ray associations in these fields, whose \HST\
morphologies and environments we analyse in the following sections.

\subsection{\textsl{HST} Observations and Reduction\label{hstobs}}
We observed three \HST/WFPC2 fields within the CDFS during the period
22--27~July 2000.  Each field received 3700s ($\sim\!2$~orbits)
exposure in the F606W filter and 5800s ($\sim\!3$~orbits) in the F814W
filter.  In each filter the exposure time was divided among four pairs
of images, arranged in a compact dither pattern with sub-pixel offsets
to ameliorate the undersampled \HST\ PSF of the
0\farcs1/pixel Wide-Field (WF) cameras.  We refer the reader to Tables 1 and 3 of
Paper II for the specific field coordinates and observation log.

After retrieving the images from the \HST\ Data Archive, we
re-calibrated them through the standard pipeline once the most
up-to-date calibration reference files for the time of the
observations became available.  We next registered each set of eight
images corresponding to a given field and filter combination via
cross-correlation, using tasks in the IRAF/STSDAS {\sl dither}
package.  To take advantage of our sub-pixel dithering, we employed
the {\sl drizzle} software (Fruchter \& Hook~2001) to combine the
images with cosmic-ray rejection onto a 0\farcs05/pixel grid (half the
WF scale) with a {\tt pixfrac} parameter of 1.0.  These drizzled
images are shown in Figure~1 of Paper~I.  The image combination is
described in more detail in Paper II.

The zeropoints of the stacked images are $ZP(F606W) = 33.322$ and $ZP(F814W) =
32.084$ in the VEGAMAG system, which we adopt throughout this paper when
quoting magnitudes or surface brightnesses.  The background noise RMS in the WF
portions of the images is $\approx\!3.1$ ADU in both F606W and F814W, and in
the PC portions is $\approx\!8.2$ ADU in F606W and $\approx\!8.6$ ADU in F814W.
For notational convenience, we refer to measurements from the F814W images as
``$I$'' and from the F606W images as ``$V$'', though we note that the F606W
passband is significantly broader and redder than Johnson $V$.

\subsection{\textsl{CXO} Observations\label{cxoobs}}
The CDFS (Giacconi et al.~2002, Rosati et al.~2002) comprises a set of
11 \CXO/ACIS-I exposures totaling $942$~ksec, taken between
October 1999 and December 2000 with varying position angle, and
centered at $\alpha=03^{\rm h}32^{\rm m}28^{\rm s}$,
$\delta=-27\arcdeg48\arcmin30\arcsec$ (J2000) in a region of low
Galactic neutral hydrogen column ($N_{\rm HI} \sim 8\times10^{19}$
cm$^{-2}$) and devoid of bright stars ($m_V \leq 14$) within
30\arcmin.  The stacked CDFS image covers $0.109$~deg$^2$ and has
$\sim\!1\arcsec$ resolution near the aimpoint, where the flux limits
reach $5.5\times10^{-17}$ \ergsscm\ in the soft band (0.5--2~keV) and
$4.5\times10^{-16}$ \ergsscm\ in the hard band (2--10~keV).  The CDFS
team has identified 346 high-significance point sources across their
stacked 1~Ms image (Giacconi et al.~2002), of which 40 fall within
the three \HST/WFPC2 fields in this study --- $\approx\!3$ times
the number of Chandra Deep Field North sources appearing in the
Hubble Deep Field--North (Brandt et al.~2001).  The overdensity of
CDFS sources in our three fields is partly by construction, as we
chose the WFPC2 aimpoints for a large overlap with the earlier
CDFS 300~ksec source list (Tozzi et al.~2001).

\subsection{WFPC2 Photometry and Source Association\label{srclist}}
Following the reduction described in Paper I, we used SExtractor
(version 2.2.1; Bertin \& Arnouts 1996) in multiple-image mode to
detect sources from the stacked $V\!+\!I$ WFPC2 images and perform
photometry of these sources on the $V$ and $I$ frames.  Table
\ref{tab:hst_cat_pap3} is the source catalog produced from the \HST\
data, containing all sources detected in the three WFPC2 fields of
view, according to the detection criteria discussed in Paper I\@.  The
catalog is complete to $I\sim26.4$ and $V\sim27.6$.  In addition to
the J2000 coordinates and $V$ and $I$ magnitudes (SExtractor
MAG\_AUTO), we include the SExtractor determinations of half-light
radius, $r_{0.5}$, and stellarity measure, $\eta$, defined such that
$\eta\to1$ for an unresolved surface-brightness profile and $\eta\to0$
for a resolved profile.  We further include each source's asymmetry
($A$) and concentration ($C$) indices, to be discussed in
\S\ref{asymconcsec}.

\placetable{tab:hst_cat_pap3}

We registered each of the three fields independently to the CDFS frame
by subtracting the error-weighted median offset of the nearest {\sl
HST}/WFPC2 optical counterpart to each {\sl CXO} source, typically
$\sim\!1\arcsec$.  After this registration, the RMS of the
optical/X-ray offsets is $\approx0\farcs5$, consistent with the {\sl
CXO} positional uncertainties (Tozzi et al.~2001).  Table
\ref{tab:hst_cxo_pap3} lists the 37 X-ray sources found by Giacconi et
al.~(2002) in the 1~Ms CDFS that were also detected in our WFPC2
observations.  We do not list the three additional Giacconi et
al.~sources in our WPFC2 fields for which we did not detect an optical
counterpart within 2\arcsec\ (see Paper II).  Table
\ref{tab:hst_cxo_pap3} includes the Giacconi et al.~catalog XID number,
IAU-format coordinate designation, hard-band (2--10~keV) X-ray flux,
and X-ray hardness ratio, defined as $(H-S)/(H+S)$ where $H$ and $S$
are the measured counts in the hard and soft (0.5--2~keV) bands,
respectively.  For convenience, we repeat from Table
\ref{tab:hst_cat_pap3} the $V$ and $I$ magnitudes and stellarities,
and the $I$-band half-light radius, concentration index, and asymmetry
index.  We also tabulate the nuclear-to-total flux ratio, to be
discussed in \S\ref{twodfitdesc}.

\placetable{tab:hst_cxo_pap3}

\section{Morphological Analyses\label{morphansec}}
We performed a number of quantitative analyses of the morphology of
the optical galaxies associated with the CDFS X-ray sources.  For the
brighter resolved sources ($I<24$, $\eta_V<0.5$), we carried out
isophotal fitting (\S\ref{isofitdesc}) and two-dimensional model fitting
(\S\ref{twodfitdesc}).  For all the WFPC2 sources, including those not
seen in X-rays, we also measured the asymmetry and concentration
indices (\S\ref{asymconcsec}), following the techniques of Conselice,
Bershady, \& Jangren~(2000) and Bershady, Jangren, \& Conselice~(2000,
hereafter BJC00).

\subsection{Isophote Fitting\label{isofitdesc}}
We fit surface brightness profiles with the IRAF isophotal analysis package
ISOPHOTE, part of STSDAS.  The package's contour fitting task {\sl ellipse}
works from an initial guess for an isophotal ellipse, then steps
logarithmically in major axis.  At each step it finds the optimal isophotal
ellipse center, ellipticity, and positional angle.  Prior to the ellipse
fitting, we use the task {\sl imedit} to mask foreground stars, neighboring
galaxies, etc., near the galaxies of interest.  Such masked pixels are ignored
by {\sl ellipse}.  We first construct the $V$-band surface brightness profile,
then apply those isophotes to the corresponding $I$-band image to obtain isophotal
colors.

Because the {\sl ellipse} algorithm averages pixels within an elliptical
annulus, it is capable of fitting isophotes out to a surface brightness
$\mu_V\approx 25.7$ mag arcsec${}^{-2}$, well below the RMS noise.  Far enough
from the galaxy center, the fitting algorithm will ultimately fail to converge,
and {\sl ellipse} enters a non-fitting mode that fixes larger ellipses to be
similar to the largest convergent isophote.  We generally ran {\sl ellipse}
non-interactively, but in cases where a peculiar galaxy surface brightness
profile sent the task into non-fitting mode prematurely, we stepped through the
isophote fitting interactively.  We show the resulting surface brightness
profiles in Figure~\ref{sbproffig}, also including the isophotal color,
position angle, and ellipticity as a function of semimajor axis.  
We have fit the one-dimensional profiles to bulge/disk models using the IRAF
task {\sl nfit1d} to assist in our classification of the optical morphologies
(Paper~II).

\placefigure{sbproffig}

\subsection{Two-dimensional Profile Fitting\label{twodfitdesc}}
We have fit parametric two-dimensional surface brightness models combining an
exponential disk, an $R^{1/4}$ bulge, and a point-source nucleus all
constrained to a common center.  The parameters of the fit are the
disk scale-length $R_d$, inclination $i_d$, and position angle ${\rm PA}_d$,
the bulge effective radius $R_b$, ellipticity $\epsilon_b$, and
position angle ${\rm PA}_b$, the bulge-to-disk flux ratio $B/D$, the
point-source to total flux ratio $P/T$, the total source flux $T\equiv
(B+D+P)$, the coordinates of the profile center, and a uniform
background level.  We pixellize the intensity distribution given by
the above 12 parameters to the same scale as the data (0\farcs05/pixel),
using a 3 times oversampled grid in the central $11\times11$ pixels
to maintain accuracy in the presence of steep intensity gradients.

We convolve this model-predicted source image by the WFPC2 PSF at the
location of the source.  Because there are too few point sources in
our images to reliably map the spatially variable WFPC2 PSF with our
data, we estimate the PSF appropriate to each \CXO\ counterpart with
the help of the Tiny Tim software (v6.0, Krist \& Hook~2001).  We
first generate a 4 times oversampled WFPC2 PSF image with Tiny Tim,
which we then convolve with the WFPC2 pixel response function
(Gaussian with $\sigma=0.28$ pixel; A. Fruchter 2003, private comm.) as
this is not automatically done by Tiny Tim for oversampled PSFs.  We
simulate the effect on the PSF of drizzling with {\sl pixfrac}=1.0
onto a $0\farcs05$/pixel grid by further convolving the oversampled
PSF image with the input ($0\farcs1$) and output ($0\farcs05$)
pixels.  Finally, we block-average the oversampled PSF up to the desired
$0\farcs05$ pixel${}^{-1}$ scale.

We fit the PSF-convolved model to an image section centered on the
source coordinates and extending well beyond the faintest detectable
isophote: as small as $64\times64$ pixels ($3\farcs2$ square) for the
$I\gtrsim22$ sources, and as large as $256\!\times\!256$ pixels
(12\farcs8 square) for the $I\approx18$ edge-on spiral CDFS
J033208.6$-$274649.  As with the isophote fitting described in
\S\ref{isofitdesc}, we mask out neighboring sources, chip edges, etc.,
and ignore such masked pixels when calculating the goodness of fit.
We derive a noise image in this region from the data itself, assuming
Poisson statistics.  We calculate a figure-of-merit
$\chi^2(\mathbf{p})$ for a given parameter set $\mathbf{p}$ by
subtracting the model image from the data, dividing this difference
image by the noise image, and summing the squares of the resulting
pixel values.  This algorithm neglects the small-scale noise
correlation introduced into the data by our drizzling procedure, but
we expect the effect to be minimal (see also Casertano et al.~2000).

We implement the parameter fitting in IDL, using as inputs: the data
image section; the associated RMS noise image; and the PSF image.  The
fitting proceeds in two stages, starting with a downhill simplex
optimization of $\chi^2(\mathbf{p})$ based on the ``amoeba''
subroutine from Numerical Recipes (Press et al.~1992).  In order to
minimize our sensitivity to starting values in the optimization, we
seed the initial simplex with randomly-generated parameter values
spanning the broadest range in the various parameters.  In the second
stage of the fitting, we try to improve upon the local minimum found
in the first stage by using the best-fit parameters as one of the
starting points for a further simulated-annealing simplex optimization
based on the ``amebsa'' subroutine from Numerical Recipes (Press et
al.~1992).

We extensively tested the fidelity of the model-fitting code with
simulated galaxy images matched to the PSF and noise properties of our
data, with representative fluxes ($10^3$--$10^5$ counts:
$19.6<I<24.6$) and scale lengths (3--10 pixels:
0\farcs15--0\farcs5), and spanning the full ranges of profile shape
parameters.  We found that the code performed very well at recovering
$P/T$ and moderately well at recovering $B/D$ for simulated galaxies
with $\gtrsim10^4$ counts ($I\lesssim22.1$).  For example, the code
typically recovered $P$ in simulated $I=22.1$ galaxies to within 10\%
over the full range of $P/T$, and for $P=0$ models it typically
returned $3\sigma$ upper limits of $P/T\lesssim 1$\% ($I\gtrsim27.1$).
For fainter simulated galaxies of $10^3$ counts ($I=24.6$), the
signal-to-noise became too poor for adequate model fits.  We therefore
restrict our discussion of the two-dimensional profile-fitting results to the $I<23$ 
galaxies in our sample, where in particular we have high confidence 
in the central point-source recovery.

To determine the uncertainty in the best-fit value for the central point-source
flux $P$, we first define the PSF-weighted chi-square of the best-fit parameter
set: $ \chi^2_{\rm PSF} \equiv \sum_{\rm pixels}{ {\rm PSF}_i}
\chi^2_i(\mathbf{p}_{\rm best})$.  Starting from the best-fit point-source flux
$P_{\rm best}$, we step away from this value with increasing $\Delta P$, at
each step carrying out a downhill simplex minimization of all the parameters
subject to the constraint $T\times(P/T)= (P_{\rm best}+\Delta P)$.  We continue
to increase $\Delta P$ in both directions until reaching the effective
$3\sigma$ perturbations, $\Delta P_{3\sigma}$, defined according to
$\chi^2_i[\mathbf{p};P=(P_{\rm best}+\Delta P_{3\sigma})] =
\chi^2_i(\mathbf{p}_{\rm best}) + 9\chi^2_{\rm PSF}$.  We have adopted this
more conservative approach to estimating the errors on the point-source
component because the $\chi^2$/pixel in the immediate vicinity of the galaxy
core is often much larger than the mean $\chi^2$/pixel across the entire
comparison region.  If this procedure indicates that $P=0$ is not excluded at
$>\!3\sigma$, we quote a $3\sigma$ upper-limit to the galaxy's point-source
component in Table \ref{tab:hst_cxo_pap3}.  Otherwise we quote $P_{\rm best}$
with its $\pm1\sigma$ errors.

\subsection{Asymmetry and Concentration Indices\label{asymconcsec}}
We measure an asymmetry index $A$ of all WFPC2 sources (both X-ray
counterparts and X-ray undetected field sources) in the three fields
following the prescription of Conselice et al.~(2000), which we
summarize here.  For each source, we determine its half-light radius
$r_{0.5}$ using SExtractor.  We then extract the square image section
$S$ centered on the source with sides of length $6\,r_{0.5}$.  We
rotate $S$ by $180\arcdeg$ to obtain the comparison image $S_{180}$.
The asymmetry index is then calculated according to the formula
$$A = \min\left({{\sum_{\rm pix}{|S - S_{180}|\Big/\sum_{\rm pix}{|S|}}}}\right) - A_0,$$
where the summation is over individual pixels, and the minimization is over a 
fractional-pixel grid (with spacing of 0.2 pixel, or 0\farcs01) of possible 
centers of rotation  .  We determine the
asymmetry zeropoint $A_0 = \min\left({\sum_{\rm
pix}{|B - B_{180}|\Big/\sum_{\rm pix}{|B|}}}\right)$ from a fixed
large square region $B$ of blank sky on the particular chip where the source
resides.  Because the background noise properties vary between pointings and
between the WFs and the PC, we cannot simply use a single $A_0$ for all sources.

Sources which happen to fall close to a chip edge, or happen to have
a spurious companion very close in projection, will have artificially inflated
asymmetry estimates in this prescription.  We have modified the asymmetry
index code (in IRAF, provided to us by C.~Conselice) to correct for the
former bias by masking the regions along chip edges where the exposure
map falls off.  When a source's image section $S$ contains such masked
pixels, we exclude those pixels (and their $180\arcdeg$-rotated
counterparts) from the summation above.  Fortunately, this chip-edge
asymmetry inflation affects only a small fraction of all sources
(e.g., two of 37 \CXO\ counterparts: CDFS XIDs 185 and 538), and should
not introduce a bias between the asymmetry distributions of \CXO\
and non-\CXO\ sources.  We do not attempt to correct the 
``spurious-companion'' bias, because we do not know {\it a priori} if 
two neighboring sources are only close because of projection.  As with the
chip-edge bias, we do not expect the spurious-companion bias to 
preferentially affect either \CXO\ or non-\CXO\ sources.
 
We list the $I$-band asymmetry indices of all field sources in
Table~\ref{tab:hst_cat_pap3}, and list the asymmetries with associated
errors (as computed by the Conselice code) for the \CXO\ counterparts in
Table \ref{tab:hst_cxo_pap3}.  We flag with colons the asymmetry
indices in Table~\ref{tab:hst_cxo_pap3} which are suspect (anomalously
large) because of substantial source overlap with a chip boundary
(CDFS XIDs 185 and 538) or likely spurious close neighbor (CDFS XID
36).  We plot the asymmetries of all sources versus magnitude in
Figure \ref{magasymfig}, denoting the \CXO\ counterparts by large
symbols keyed to morphological type as classified in Paper~II: E/S0 (circles);
S/Irr (squares); unresolved (stars); and indeterminate (triangles).
We also plot the median $A$ and median uncertainty in $A$ for field
sources in successive 1-mag bins (crosses with error bars).  We see no
significant trend in $A$ with magnitude, but the typical measurement error
becomes substantial ($\gtrsim 0.1$) by $I\approx23$.  We therefore
restrict our samples to $I<23$ when discussing the implications of the
asymmetry measurements in \S\ref{asymdisc}.

\placefigure{magasymfig}

We measure the concentration index $C$ of all sources in the three fields
following the prescription of BJC00.  Using SExtractor to measure the radii
containing 20\% and 80\% of each source's flux ($r_{0.2}$ and $r_{0.8}$,
respectively), we then compute the concentration index as $C \equiv
5\log(r_{0.8}/r_{0.2})$.  Given this definition, a pure exponential profile has
$C=2.80$, while a pure $R^{1/4}$ profile has $C=5.27$.  A Gaussian profile has
$C=2.15$, and thus an observed PSF-convolved exponential or $R^{1/4}$ profile
will appear less concentrated than the theoretical values.  We list the
concentration indices for the individual \CXO\ counterparts in Table
\ref{tab:hst_cxo_pap3} and plot the distribution of concentration with magnitude
in Figure \ref{concmagifig} for both the X-ray counterparts (large symbols) and
all the sources in the field (small symbols).  We note that the high-asymmetry
galaxy CDFSJ033211.0$-$274343 ($A=0.47$), which is significantly truncated by a
chip edge, is the lone low-concentration outlier among the brighter
\CXO-detected sources.

Because the SExtractor detection algorithm requires a minimum number
of connected pixels above a minimum counts/pixel threshold, sources at
the faintest flux levels will be preferentially detected if they have
a flatter intensity profile.  This detection bias should be reflected
in a systematic decline of the mean $C$ at the faintest flux levels,
which we indeed observe in Figure \ref{concmagifig} at
$I\gtrsim23$--24.  When discussing the implications of the
concentration measurements in \S\ref{concdisc}, we avoid the
complications of this detection bias by restricting our analysis to
the $I<24$ population.

We estimate the error on $C$ because of uncertainty in the total flux of
the source by measuring the additional two flux-radius ratios $C_\pm
\equiv 5\log(r_{0.8(1\pm\sigma_f)}/r_{0.2(1\pm\sigma_f)})$, where
$\sigma_f$ is the fractional error on the total flux.  The total error
on $C$ should also include the uncertainty in the two isophotal radii
values because of the pixel RMS noise.  This error on the flux radii is
not reported by SExtractor, and is cumbersome to compute for all
sources.  However we do not expect this additional uncertainty to
greatly exceed our estimate based on $C_\pm$.  In Figure
\ref{concmagifig} we plot the median $C$ and median uncertainty in $C$
for field sources in successive 1-mag bins (crosses with error bars).
The uncertainty in $C$ is $\lesssim0.05$ for sources with
$I\lesssim24$.

\placefigure{concmagifig}

\section{Discussion\label{discsec}}
\subsection{Surface Brightness Profile Results for the \textsl{CXO} Counterparts
\label{sbdisc}}
Examining the isophotal fitting results for the 22 resolved $I<24$ \CXO\
counterparts (Fig.~1), we find that about half of the objects have a relatively
flat color profile all the way into the nucleus, while a slightly smaller
fraction display a nucleus that is at least half a magnitude bluer at its core
than the average color across the rest of the galaxy.  Three objects
display nuclei that are redder by at least half a magnitude than
their average global color.

We have searched for correlations between color gradient results and the X-ray
properties of the galaxies.  The three redder-centered objects appear to have
somewhat harder X-ray spectra, while the flat-color objects are softer and in
particular include all objects detected solely in the \CXO\ soft band
(0.5--2~keV).  The bluer-centered objects have hardness ratios that are
intermediate in value, covering the entire range from soft to hard.  There is
also some suggestion that the bluer-centered objects have a slightly higher
$L_X$ on average than the flat-color and redder-centered objects.

The correlation between higher $L_X$ and blue central color gradient
is expected since the bluer-centered objects are probably all type~1
AGN and are therefore detected at somewhat higher distances on
average.  This is confirmed by the optical spectroscopy that we have
so far: all 3 of the 8 bluer-centered objects for which we have
spectroscopic classifications (Szokoly et al.~2003, in prep.)  are
type~1's; all 4 of the spectroscopic classifications available for the
16 flat-color and redder-centered objects suggest that they are type~2's.
Finally, the bluer-centered objects are found mostly in elliptical
hosts (according to our morphological classifications), while the
flat-centered objects are found in both spiral and elliptical hosts.

\placefigure{ptsrcfig}

\subsection{The Ratio of Obscured to Unobscured AGN at $z\sim 0.5$--1
\label{ptsrcdisc}}
Most of the optically brighter \CXO\ counterparts for which we
estimated the nuclear point-source flux (\ref{twodfitdesc}) also
have measured redshifts (Szokoly et al., in prep.) which imply $L_X
\gtrsim 10^{42.5}$.  Such large X-ray luminosities are associated with
AGN or intense starbursts --- in either case the X-ray output of the
host galaxy is dominated by the nuclear regions.  Thus we can use our
estimates of the unresolved nuclear flux in the CDFS counterparts to
provide a much cleaner investigation of X-ray sources' $F_X/F_{\rm
opt}$.  Figure \ref{ptsrcfig} shows two plots of 2--10~keV $F_X$
versus~$F_{\rm opt}$ for the 1~Ms CDFS sources in our fields: the top
panel shows $F_{\rm opt,tot}$ of the entire host (including nucleus), analogous to
Figure~8 of Paper~I for the 300~ksec CDFS sources; the bottom panel
shows $F_{\rm opt,nuc}$ of the counterparts which are either unresolved (filled star
symbols) or resolved but with finite unresolved nuclear flux (open
star symbols).  In the bottom panel, we also pair each $F_{\rm
opt,nuc}$ with the corresponding $F_{\rm opt,tot}$ (dotted symbols)
for comparison with the top panel.  Here and throughout the
rest of the discussion, we focus on the $I$-band data in preference to
the $V$-band, since only the former samples rest-frame optical light
($\gtrsim\!4000$\AA) at the substantial redshifts ($z\sim0.5$--1) of
even the brighter resolved \CXO\ counterparts.

It is immediately obvious from Figure \ref{ptsrcfig} that the scatter
in $F_X/F_{\rm opt}$ is greatly reduced when considering only the
optical flux originating from an unresolved nucleus --- most have
$\log(F_X/F_{\rm opt,nuc}) \sim 1.1\pm0.3$.  In particular, we see
that the $F_X/F_{\rm opt,nuc}$ values for the resolved CDFS sources
are now consistent with the narrow range of $F_X/F_{\rm opt}$ of the
several \HST-unresolved counterparts --- all
spectroscopically-confirmed type~1 AGN (Szokoly et al., in prep.).  We
may therefore infer that we are seeing type~1 AGN in these additional
resolved hosts, bringing the total fraction of brighter CDFS
counterparts with unobscured AGN to $\sim\!50$\%.  

Even these brighter optical CDFS counterparts are at substantial
distances ($z\sim 0.5$--1; Szokoly et al., in prep.), yet their type~1
fraction is comparable with recent measures of unbiased AGN samples at
lower redshift (Wilkes et al.~2002).  The hard-X-ray sensitivity of
\CXO\ can detect even highly-obscured AGN ($N_H \sim
10^{24}$~cm${}^{-2}$) of moderate luminosities ($L_X \gtrsim
10^{42.5}$ erg s${}^{-1}$) in the 1~Ms CDFS out to $z\gtrsim1$,
leading us to conclude that our current sample, albeit modest, does
not support a steep redshift evolution in the type~2/type~1 ratio as
favored by recent AGN synthesis models (e.g.,~Gilli et al.~2001).

\subsection{Asymmetry of X-ray Hosts versus~Field Galaxies
\label{asymdisc}}
Having measured the asymmetry and concentration indices of all sources
appearing in our three WFPC2 pointings (\S\ref{asymconcsec}), we can make
statistical comparisons of these morphological indicators between the 
\CXO-detected and the \CXO-undetected sources.  Figure \ref{asymcdffig}
shows the cumulative distribution of asymmetry index $A$ for the 21 
\CXO-detected and \HST-resolved sources with $I<23$ (heavy solid line)
and for the remaining 267 galaxies with $I<23$ but undetected by \CXO.
The two distributions are statistically indistinguishable according to the
Kolmogorov-Smirnov (K-S) test, which gives a 56\% probability of the null
hypothesis that they are drawn from the same underlying distribution.

\placefigure{asymcdffig}

Our finding that the brighter resolved CDFS sources, almost all at moderate
redshifts $z\sim0.5$--1, have a strikingly similar asymmetry
distribution to the $I<23$ field may be compared with earlier morphological
studies of the {\sl local} AGN population.  For example, Corbin~(2000) finds
that the 45 galaxies in the Seyfert and LINER subsets of the BJC00 sample do
not show significant asymmetry index differences compared with the 60 galaxies
among the non-active subsets.  $N$-body simulations of galaxy interactions
(Walker, Mihos \& Hernquist 1996) suggest that even minor mergers induce
sufficient morphological disturbances to be detectable via the asymmetry index
up to $\sim\!1$~Gyr after the onset of the merger.  On this basis,
Corbin~(2000) concludes that minor mergers either have {\em no} role in the
triggering of lower-luminosity AGN, or that the AGN manifests only in the late
stages of the mergers --- implying that the nuclei in his nearby sample have
only been active within the last $\sim\!0.1$~Gyr.

Inasmuch as the hard X-ray sensitivity of the CDFS provides a
near-complete census of both obscured and unobscured AGN at
$z\lesssim1$ down to modest luminosities, similar to the BJC00
sample, the CDFS counterparts' unexceptional asymmetries and paucity
of obvious merger candidates ($A>0.35$; C.~Conselice 2003, private comm.)
now suggest that very recent merger history is not a good indicator of
AGN activity over a substantial range of lookback time.  If galaxy
mergers are indeed a primary driver of AGN fueling, as has been
commonly hypothesized (e.g.~Gunn~1979, Dahari~1984, Roos~1985,
Taniguchi~1999), then the epoch of AGN activity cannot closely
coincide with that of merger-induced morphological disturbance ---
$\lesssim\!10^9$ yr.

\subsection{Companions to X-ray Hosts versus~Field Galaxies}

As a further check into the merger--AGN connection, we took a census
of near neighbors to the CDFS counterparts.  In Figure \ref{nnfig} we
plot a histogram of the number of $I<23$ galaxies appearing within
8\arcsec\ of the $I<23$ CDFS counterparts (heavier solid line).  For
comparison, we also plot the histogram of $I<23$ galaxies appearing
within 8\arcsec\ of the \CXO-undetected $I<23$ population (lighter
solid line).  We adopt an angular separation threshold rather than a
projected linear distance because we currently lack redshift
information for most of the field galaxy control sample.  However we
note that 8\arcsec\ corresponds to a narrow range of angular diameter
distance, $\approx50$--65~kpc, over a broad redshift range $0.5<z<5$
in our adopted cosmology.  This redshift range includes almost all the
CDFS counterparts at $I<23$, whose redshift survey is virtually
complete (Szokoly et al., in prep.), and presumably the majority of
the $I<23$ field galaxies.  If the $I<23$ galaxies were not clustered,
an 8\arcsec\ radius would on average contain 1.1 such objects given
their source density over the $\approx\!14.7$~arcmin${}^2$ of our
three WFPC2 fields.  We show the corresponding Poisson distribution in
Figure \ref{nnfig} (dotted line) for reference.

Compared with the X-ray-undetected $I<23$ population, the CDFS
counterparts show an enhancement in the fraction with a single near
neighbor and a corresponding deficit of isolated systems; the fraction
with multiple companions is almost identical between the two
populations.  While it is obvious that the near-neighbor frequency
histogram (Fig.~\ref{nnfig}) shows greater discrepancy between the
X-ray host and field populations than the asymmetry parameter,
nonetheless the $\chi^2$ statistic between the two histograms gives
only a modest rejection of the null hypothesis (Prob$(\chi^2)=0.019$)
that the nearest-neighbor frequencies are consistent.  Larger samples
are clearly needed to establish the significance of a bias against
isolated CDFS sources (see \S\ref{concsec}).  As the current data do
not support a significant enhancement in the fraction of CDFS sources
with near neighbors, this reinforces our interpretation from the
asymmetry analysis (\S\ref{asymdisc}) that there is no strong
association between low- to moderate-luminosity AGN at moderate
redshifts and recent/ongoing merger activity of the host galaxies.
Similar conclusions at low redshift have emerged from recent studies
of the AGN fractions among close pairs versus the field (Barton,
Geller, \& Kenyon~2000), and of the companion fractions of AGN versus
non-AGN (Schmitt~2001).

\placefigure{nnfig}

\subsection{Central Concentration of X-ray Hosts versus~Field Galaxies
\label{concdisc}}

Figure \ref{conccdffig} shows the cumulative distribution of concentration
index $C$ for the 21 \CXO-detected and \HST-resolved sources with
$I<23$ (heavier line) and for the remaining 267 galaxies with $I<23$ but
undetected by \CXO\ (lighter line).  In marked contrast to the result for the
asymmetry index, the two distributions of $C$ are clearly distinct and have a
low K-S probability of $4.3\times10^{-6}$ for the null hypothesis.

\placefigure{conccdffig}

Even though we have restricted this comparison to the {\sl
HST}-resolved CDFS counterparts, i.e.~those whose optical flux is not
dominated by the central engine, one may still worry that their
concentration indices may be biased by the presence of optical flux
from the AGN.  Indeed we have found several of the CDFS host galaxies
to be best-fit with a nuclear point source comprising $\sim\!1-10$\%
of the total flux (Fig.~\ref{ptsrcfig}).

In theory, our two-dimensional modeling could be used to refine the $C$ estimates
by removing nuclear point-source bias.  For example, one may consider
computing $C$ from a galaxy image {\sl after} subtracting the best-fit
nuclear point source.  The model PSF does not precisely match the
shape of the true PSF, however, and scaled PSF subtraction leaves
relatively large-amplitude residuals in the host's central pixels
which hamper $C$ estimation via SExtractor.  Alternately, one may
consider computing $C$ from a model galaxy image constructed from the
other, non-point-source, parameters in the two-dimensional fit.  In practice this
approach also fails, as our simulations indicate the two-dimensional fitting
robustly recovers the point-source flux but not the input bulge and
disk shape parameters.  Furthermore, a proper comparison with the
field control sample in either case would require extending the two-dimensional
modeling and ``nuclear point-source correction'' to the whole
population, which is beyond the scope of the current study.
Fortunately, our calculations below suggest that any enhancement of
$C$ from nuclear point-source flux is at most a small fraction of the
observed $C$ discrepancy (Fig.~\ref{conccdffig}) between \CXO\ hosts
and field.

The effect of added point-source flux to a galaxy's concentration
index is not straightforward, since the near-Gaussian \HST\
PSF is significantly less concentrated than
an exponential or $R^{1/4}$ profile.  Small additions of nuclear flux
will cause $C$ to rise as $r_{20}$ shrinks faster than $r_{80}$, but
when the point-source becomes sufficiently dominant the concentration
will peak and then collapse to the lower $C$ of the PSF.  We have
attempted to quantify this effect by creating a suite of model galaxies
based on a fiducial galaxy profile typical of our $I<23$ CDFS
counterparts: a PSF-convolved, $I=22$ deVaucouleurs profile with $R_e
= 5$~pixels (0.25\arcsec).  We then steadily increase the flux fraction
in the point-source nucleus, and measure $C$ as a function of the
point-source fraction.

The result is plotted in Figure \ref{concbiasplot}, where we see that
the peak increase of $C$ is only $\sim\!0.1$.  Over the range
corresponding to our measured point-source fractions, the deviation is
$<\!0.05$.  This is only $\approx\!10\%$ of the discrepancy seen in
Figure~\ref{conccdffig}, from which we conclude that the AGN host
concentration bias is truly related to differences in the host galaxy
structure, and not to the possibility of nuclear point-source flux
contribution.

\placefigure{concbiasplot}

Graham et al.~(2001) have found a tight correlation between
SBH mass and host galaxy concentration index
exists among the nearby galaxy samples which previously established
the close correlation between SBH mass and host bulge velocity
dispersion (Ferrarese \& Merritt~2000, Gebhardt et al.~2000).
Although our initial analysis reveals little correlation between $L_X$
(as a proxy for SBH mass) and $C$ for the $I<23$ CDFS host galaxies,
it is likely that our sample --- almost all with hard X-ray
luminosities indicative of AGN ($L_X > 10^{42.5}$ ergs s${}^{-1}$) ---
harbor SBHs at the high end of the SBH mass function relative to the
X-ray undetected galaxies in our fields.  Our WFPC2 findings therefore
suggest, for the first time, that the close linkage between SBH mass
and host galaxy properties extends to the substantial look-back times
($z\sim0.5-1$) typical of the $I<23$ CDFS X-ray sources.

\section{Summary\label{concsec}}
The deep fields observed with the {\sl Chandra X-ray Observatory} have
produced the first large uniform catalogs of X-ray sources down to
modest AGN luminosities ($L_X\sim10^{42-42.5}$~erg~s${}^{-1}$) at
substantial redshift.  Given the good hard-X-ray sensitivity of the
CDFS observations, this catalog is expected to provide a near-complete
census of both obscured and unobscured AGN at $z\lesssim1$, whose
redshifts appear to be concentrated in the range $z\sim0.5-1$ (Rosati
et al.~2002; Szokoly et al., in prep.).  We have presented this first
quantitative morphological study of many optical counterparts of these
distant AGN based on our deep \HST\ observations of three WFPC2 fields
in the CDFS.

We have compared the asymmetry indices of $I<23$ AGN hosts with the
much larger population of similarly bright X-ray-undetected field
galaxies.  We find a very good agreement between the asymmetry
distributions of the two populations (Fig.~\ref{asymcdffig}).
Furthermore, the \CXO\ counterparts show only a modest difference from
the field population in the frequency of nearby companions
(Fig.~\ref{nnfig}).  These findings are similar to studies of
comparable low-redshift populations, and now suggest that recent
merger history is not a good indicator of AGN activity over a
substantial range of lookback time.

Our PSF-convolved two-dimensional profile fitting of the \HST-resolved \CXO\
counterparts shows that over a third have significant nuclear
point-source emission.  The $F_X/F_{\rm opt}$ of these unresolved
nuclear sources is consistent with the narrow $F_X/F_{\rm opt}$ range
of the several \HST-unresolved counterparts (Fig.~\ref{conccdffig}).
We therefore infer that the total fraction of $I<24$ CDFS counterparts
harboring type~1 AGN is $\sim\!50$\%, consistent with recent findings
for unbiased AGN samples at lower redshift (Wilkes et al.~2002).  If
the large type~1 fraction inferred from our modest sample were
representative of the overall $z\sim 0.5$--1 AGN population, this
would be problematic for recent AGN synthesis models (e.g.~Gilli et
al.~2001) which favor a sharp rise in the type~2/type~1 ratio out to
$z\sim1$.

Our comparison of the concentration indices of \CXO\ counterparts with
the field reveals a clear bias toward higher $C$ among the \CXO\ host
galaxies (Fig.~\ref{conccdffig}).  This $\sim\!5\sigma$ result
does not appear to be an artifact of heightened optical flux from the
CDFS hosts' nuclei, and may represent the first evidence that the
locally-observed correlation between SBH mass and host-galaxy
properties, including concentration index, is already in place by
$z\sim 1$.

To press upon these results regarding the merger--AGN connection and
SBH--host-galaxy connection at moderate- to high-redshift, we are
undertaking extensive \HST/Advanced Camera for Surveys (ACS)
multicolor imaging of the two {\sl Chandra} Deepest Fields as part of
the Great Observatories Origins Deep Survey (GOODS, Dickinson \&
Giavalisco~2003).  The GOODS data will both increase by
$\gtrsim\!10$ times the available number of \HST-imaged AGN hosts
and comparison field galaxies in \CXO\ $\gtrsim\!1$~Ms fields and also
will provide extensive spectroscopic and photometric redshift coverage to
complement the ACS morphologies.  With $>\!400$ \HST\ counterparts of
$\gtrsim1$~Ms \CXO\ sources, we will solidly verify the morphological
and environmental conclusions we have drawn from the 37 CDFS host
galaxies in the present work.  Furthermore, the large sample in
combination with multicolor imaging and redshift information will
allow us to track the {\em evolution} of AGN versus~non-AGN environment
and rest-frame $B$ morphology out to $z\sim1.3$ (Grogin et al.~2003).

\acknowledgements 
We thank C.~Conselice for insightful discussions and
for graciously providing his asymmetry-index code for our use.  We
thank R.~Gilli and the anonymous referee for comments which
strengthened the manuscript.  We gratefully acknowledge the award of
\HST\ Director's Discretionary time in support of this project.  We
also acknowledge support for this work which was provided by NASA
through GO grants GO-08809.01-A and GO-07267.01-A from the Space
Telescope Science Institute, which is operated by AURA, Inc., under
NASA Contract \hbox{NAS 5-26555}.

\clearpage
\newcommand{\subplotwid}{2.1in}
\begin{figure}[bpt]
\epsfig{figure=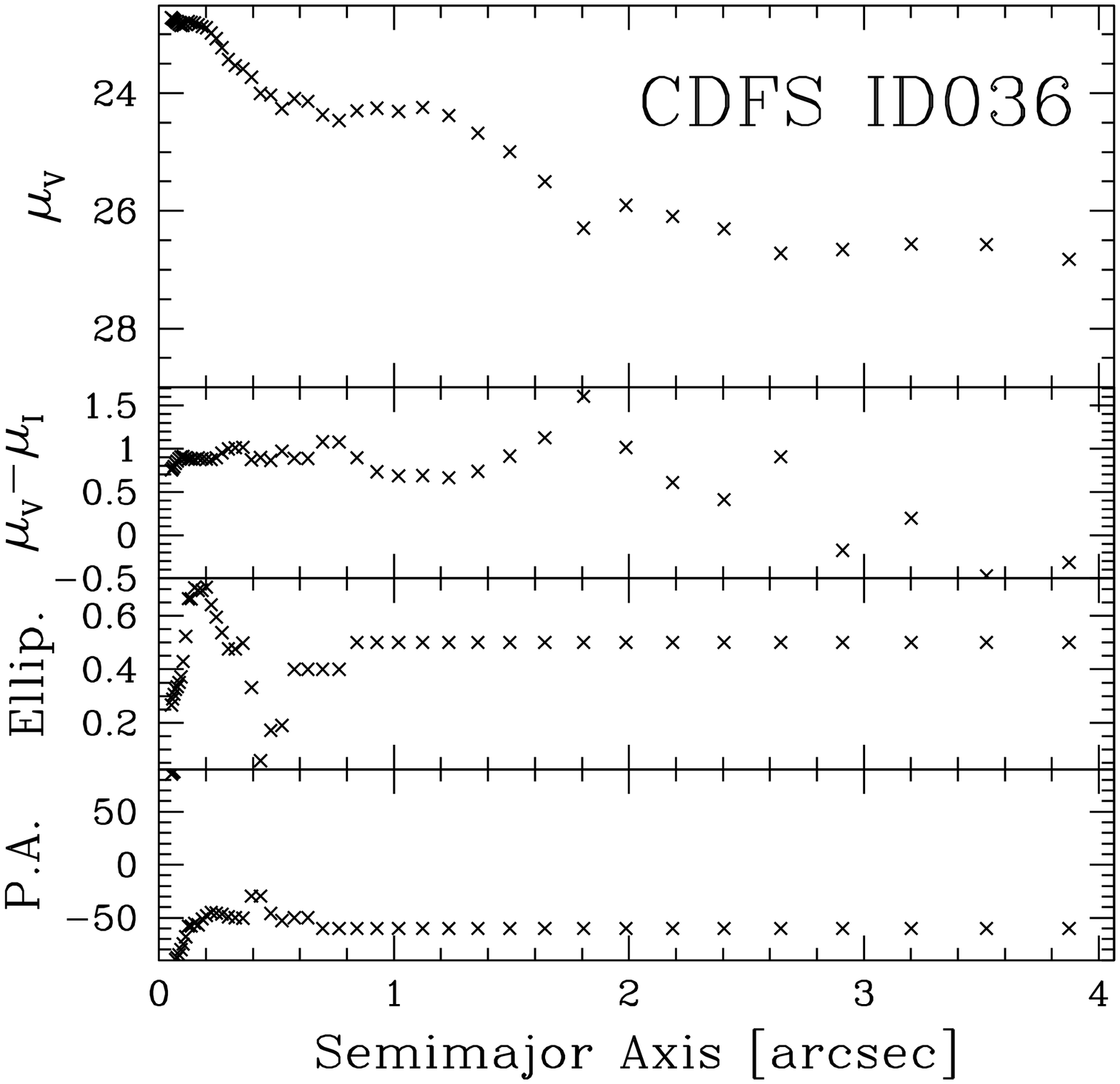,angle=0,width=\subplotwid}
\epsfig{figure=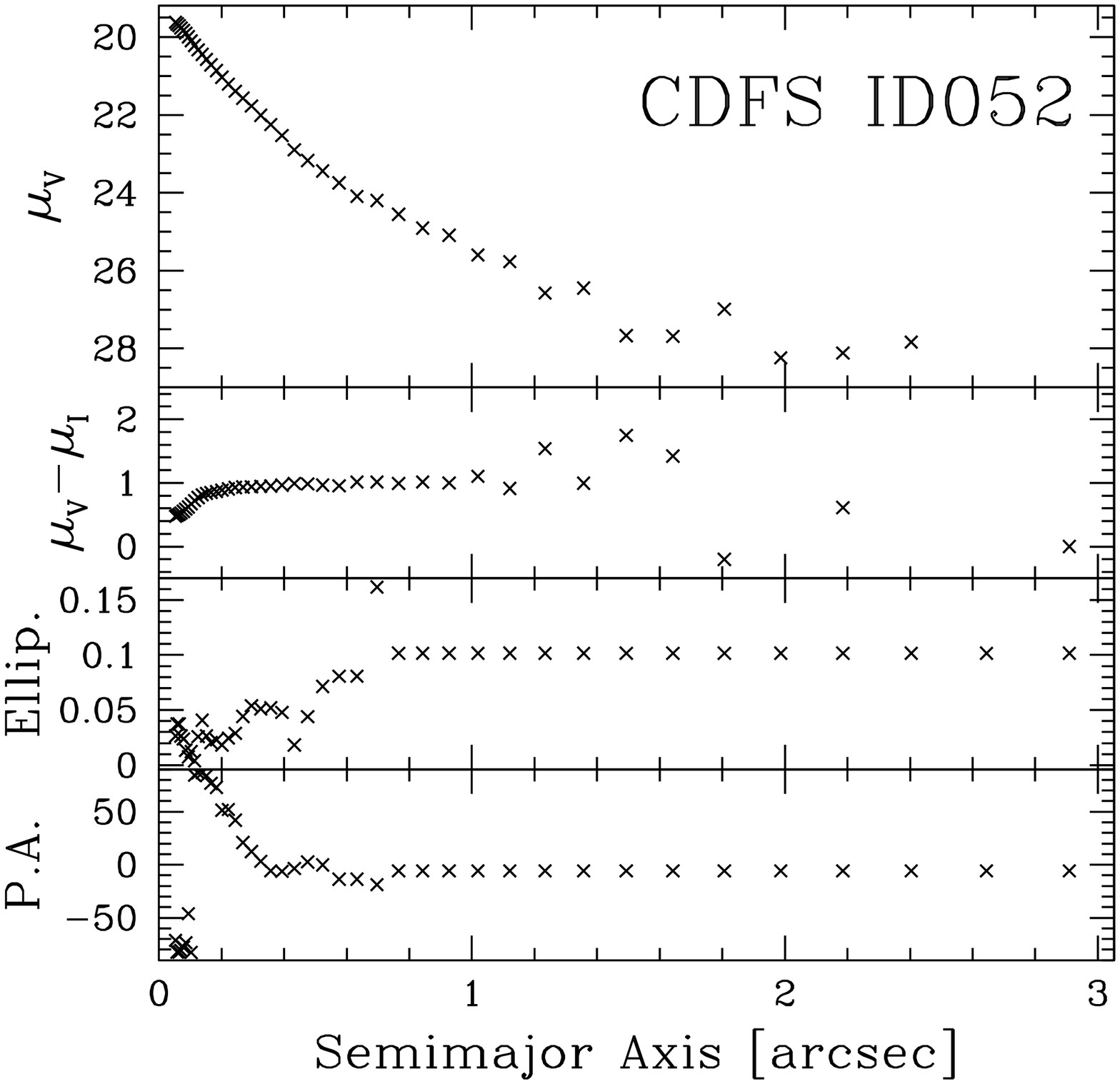,angle=0,width=\subplotwid}
\epsfig{figure=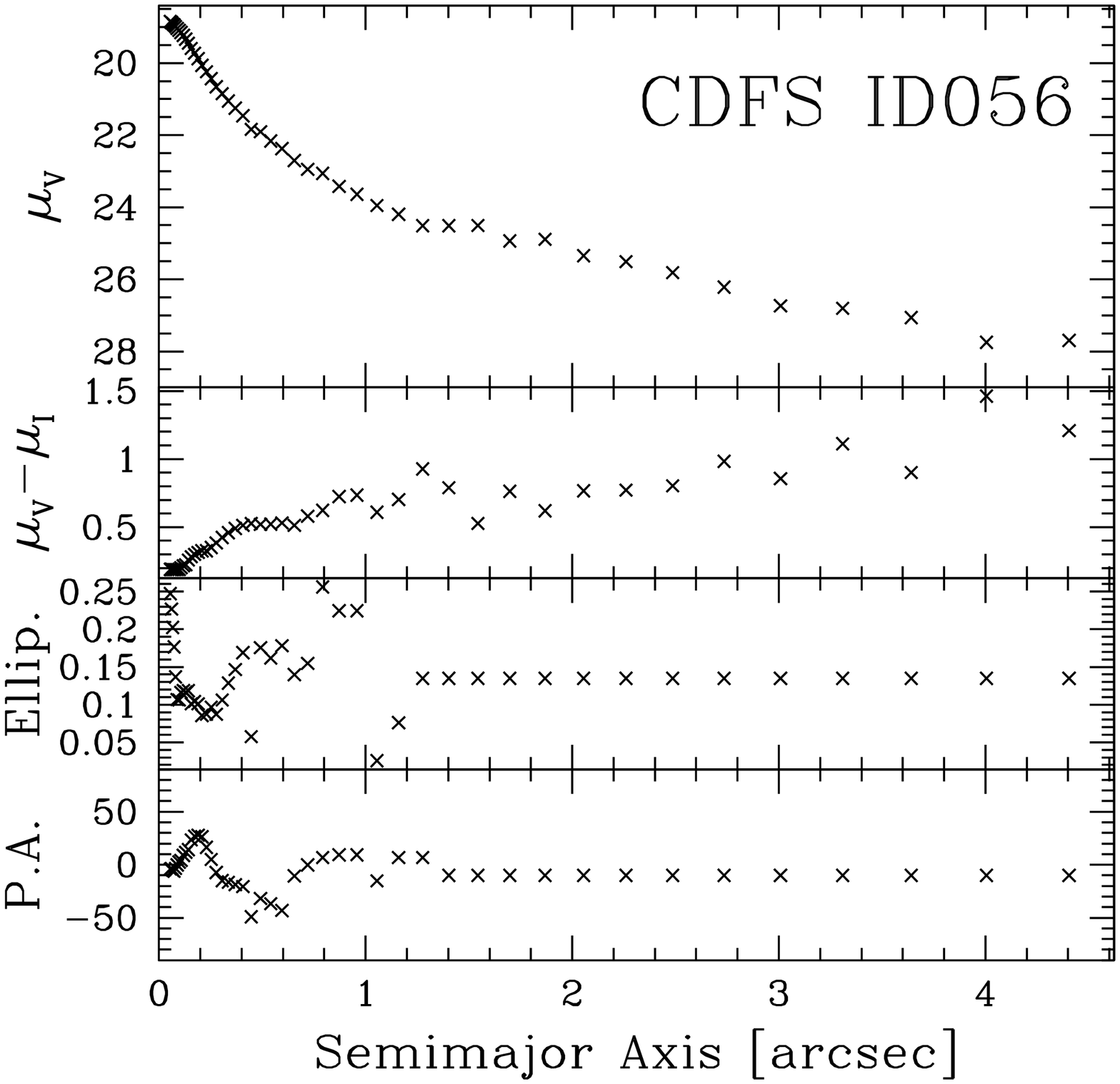,angle=0,width=\subplotwid}\\
\epsfig{figure=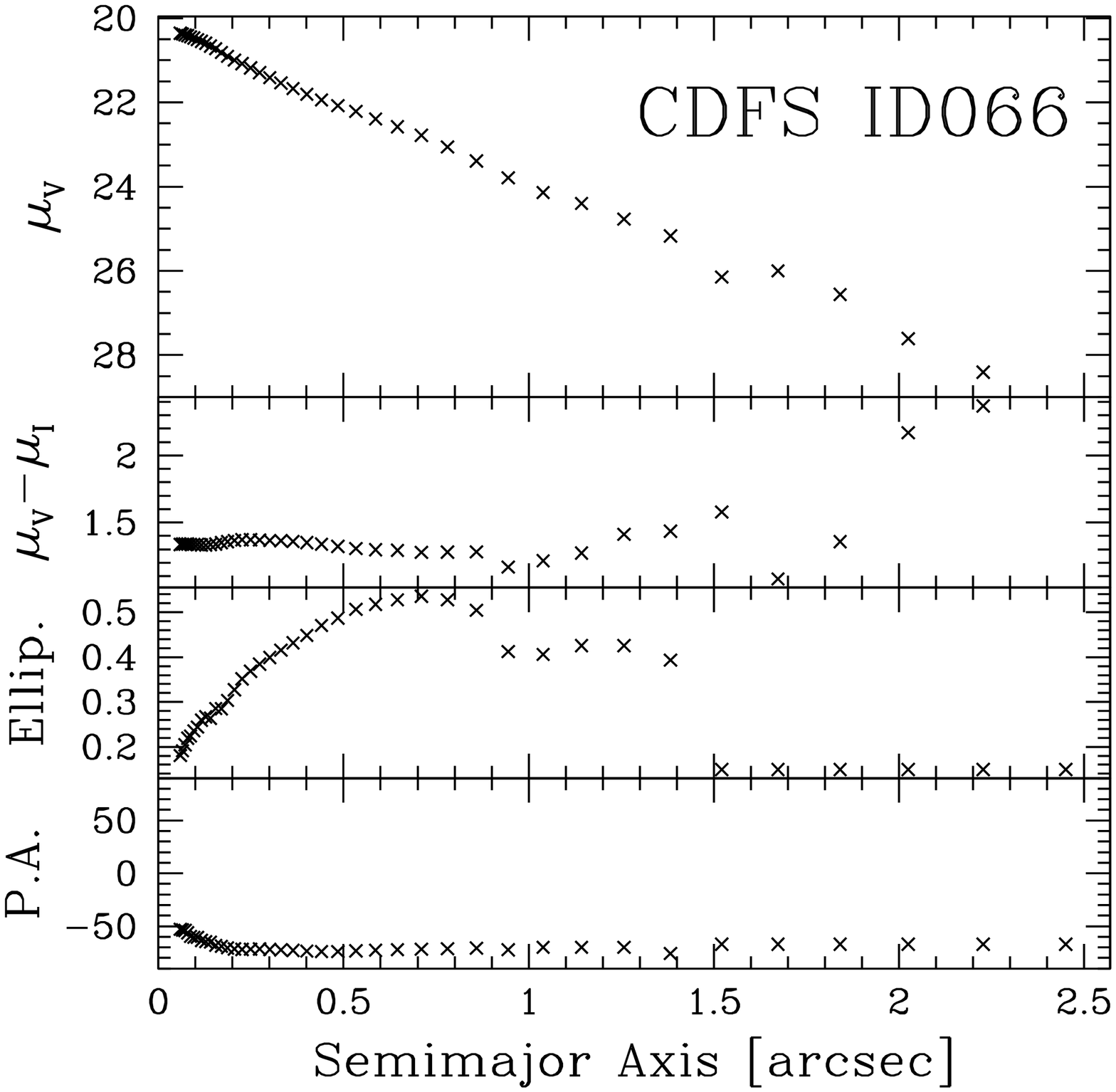,angle=0,width=\subplotwid}
\epsfig{figure=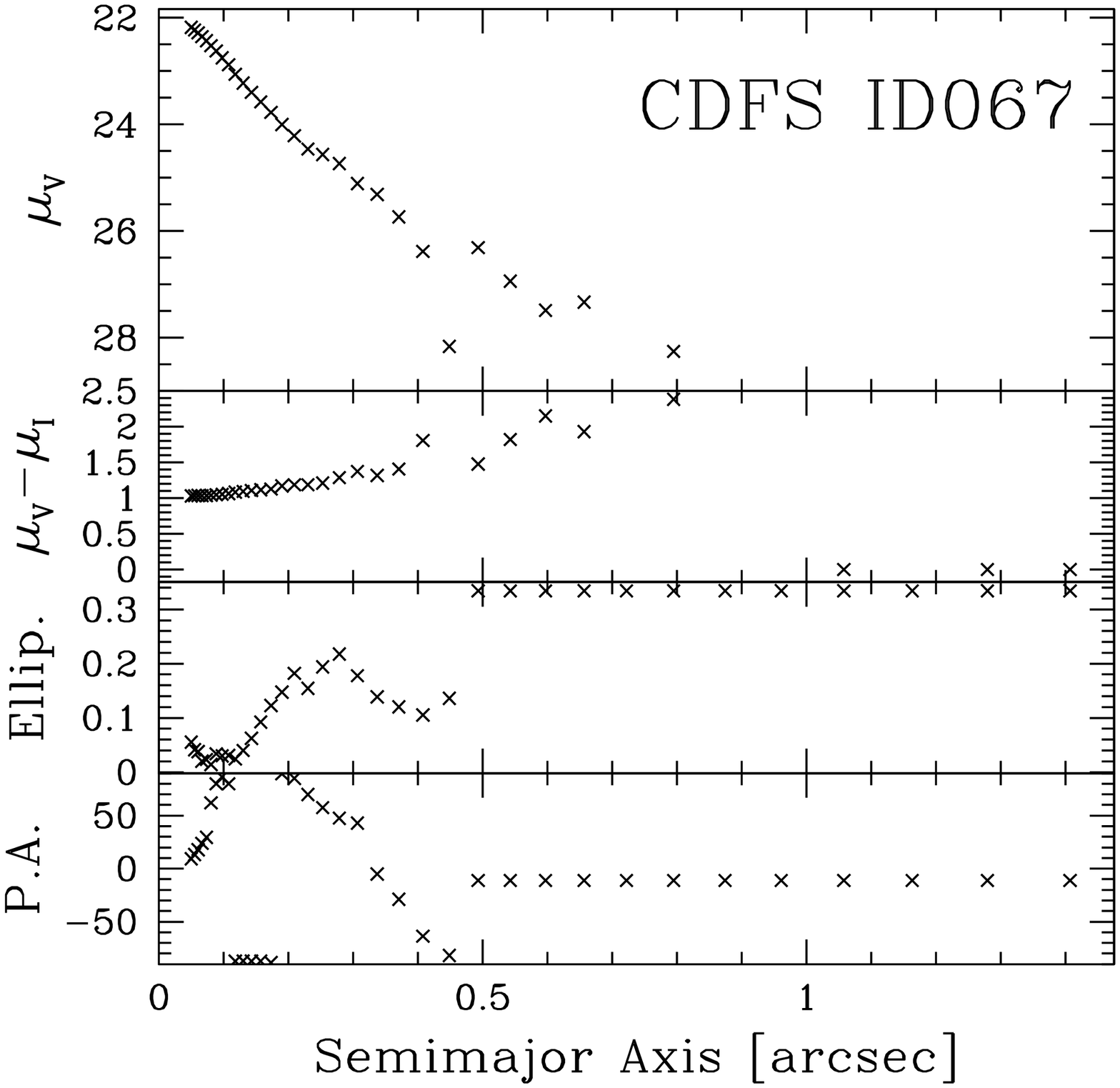,angle=0,width=\subplotwid}
\epsfig{figure=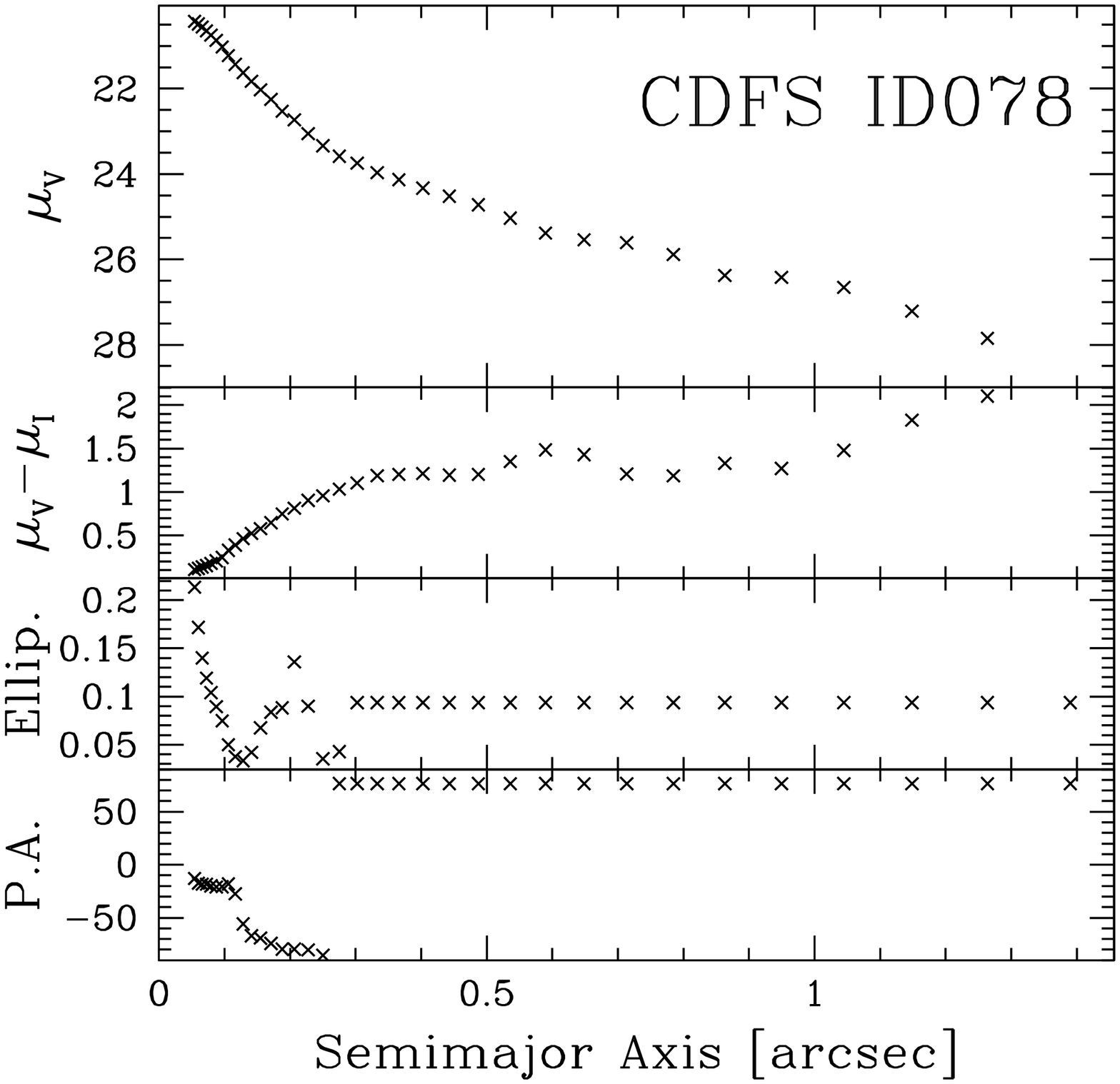,angle=0,width=\subplotwid}\\
\epsfig{figure=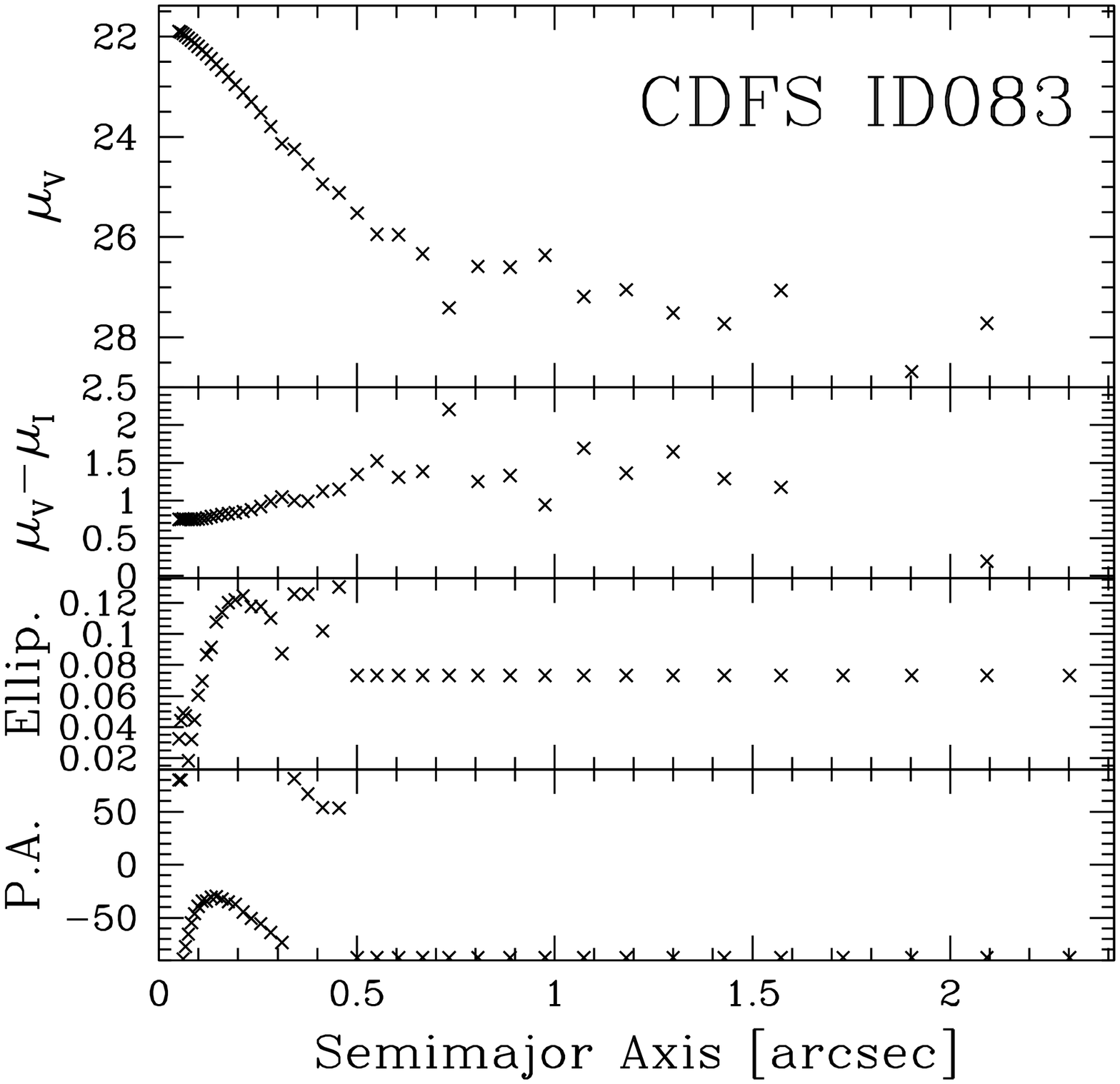,angle=0,width=\subplotwid}
\epsfig{figure=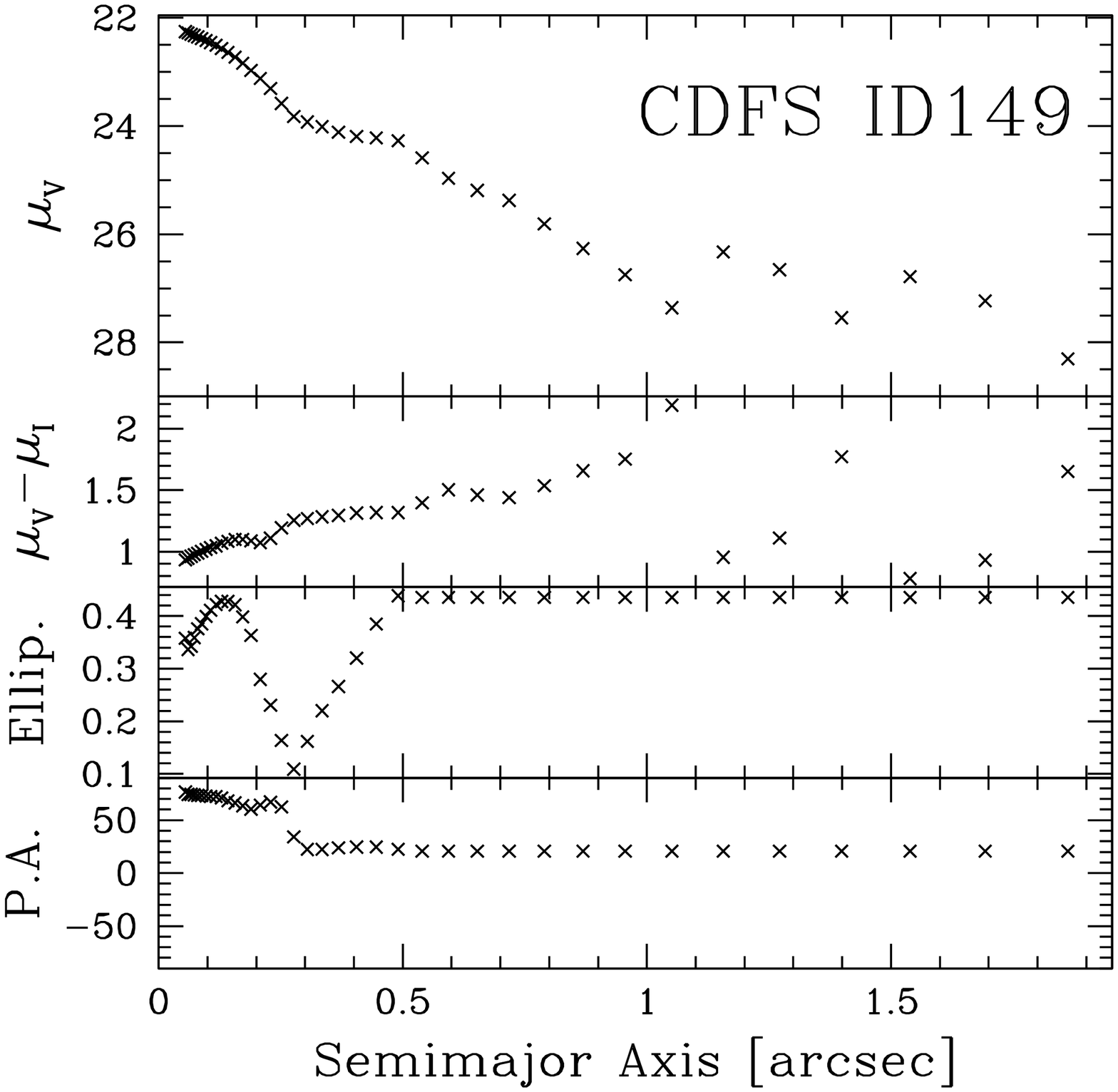,angle=0,width=\subplotwid}
\epsfig{figure=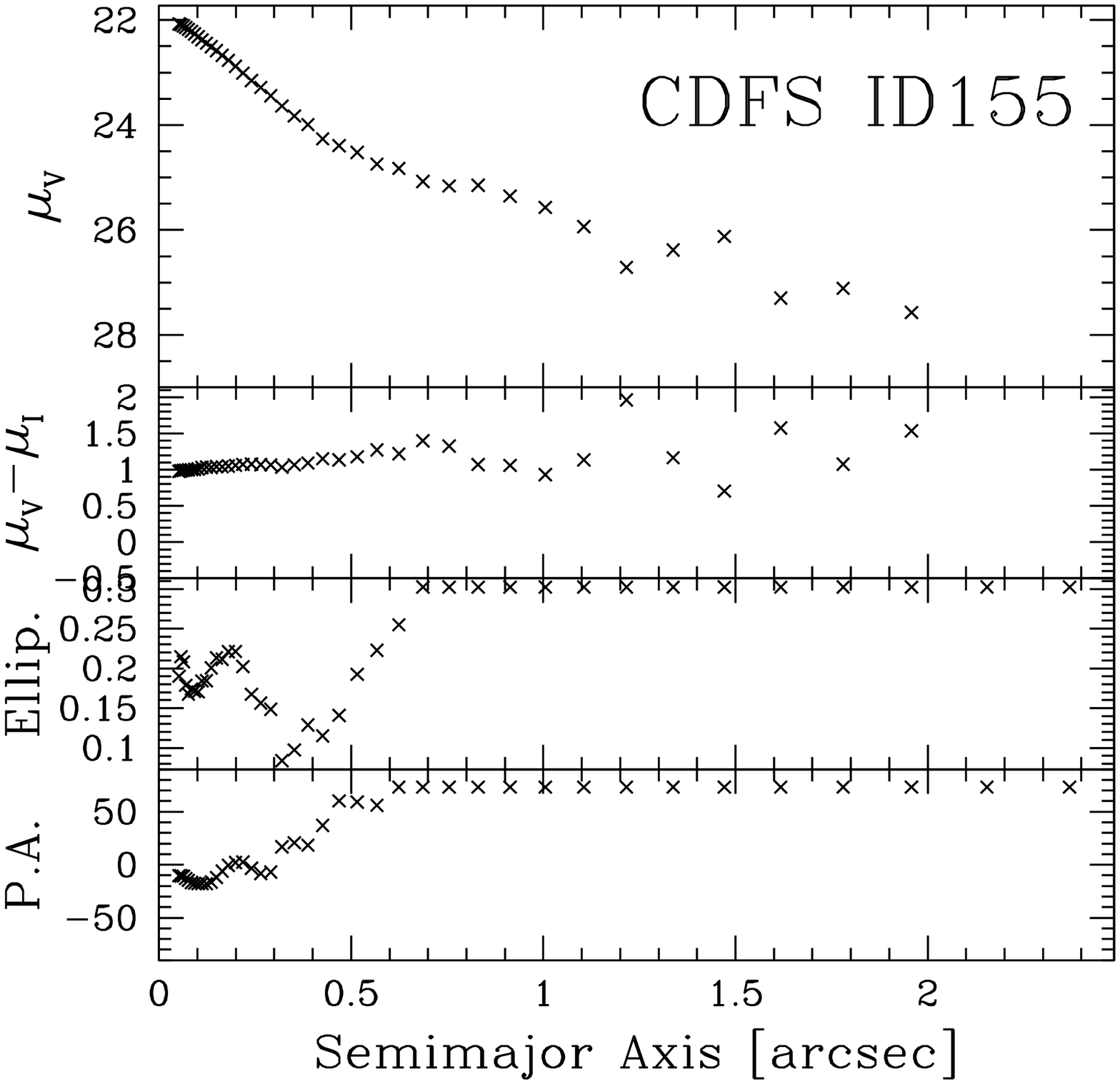,angle=0,width=\subplotwid}\\
\epsfig{figure=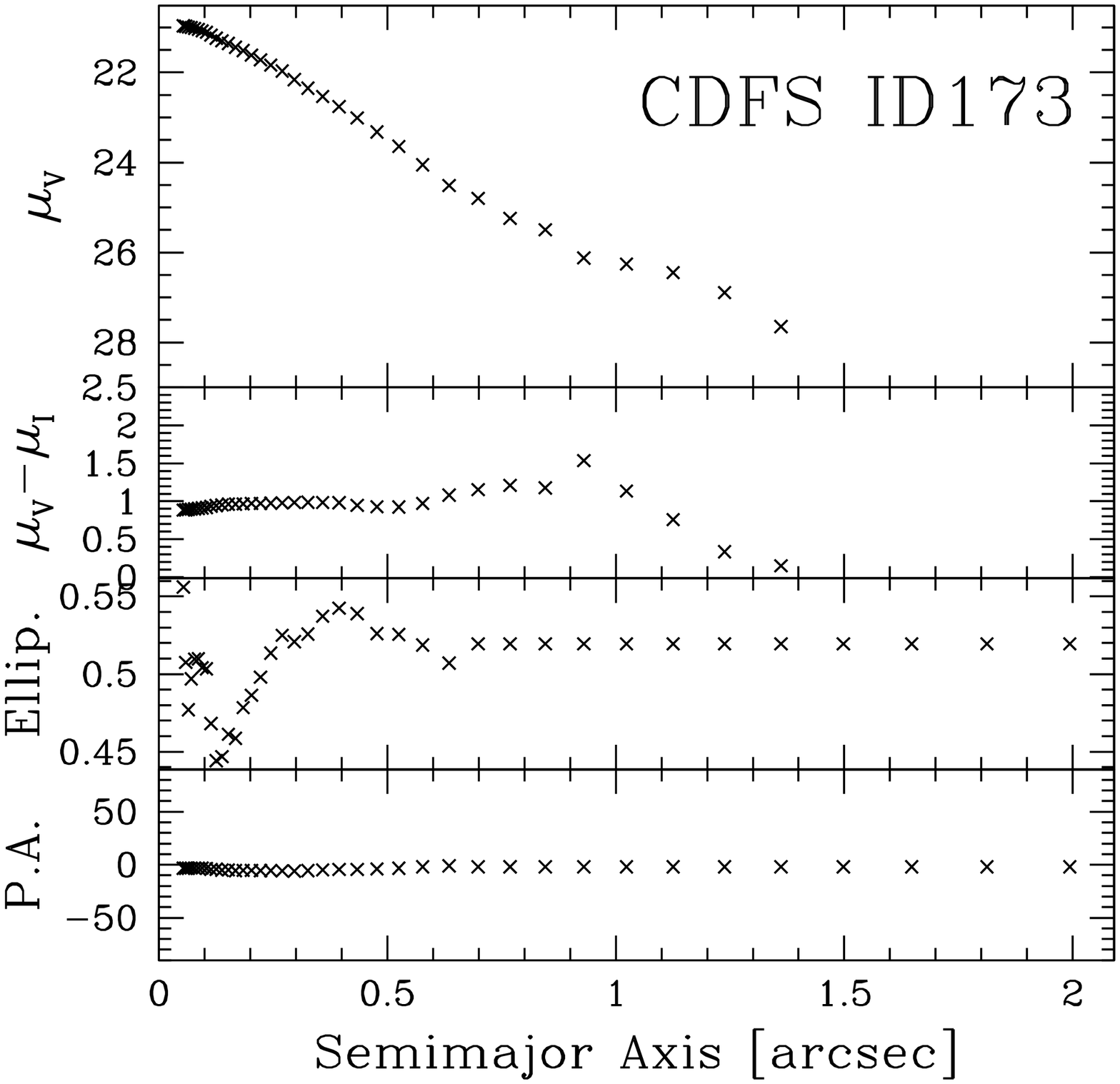,angle=0,width=\subplotwid}
\epsfig{figure=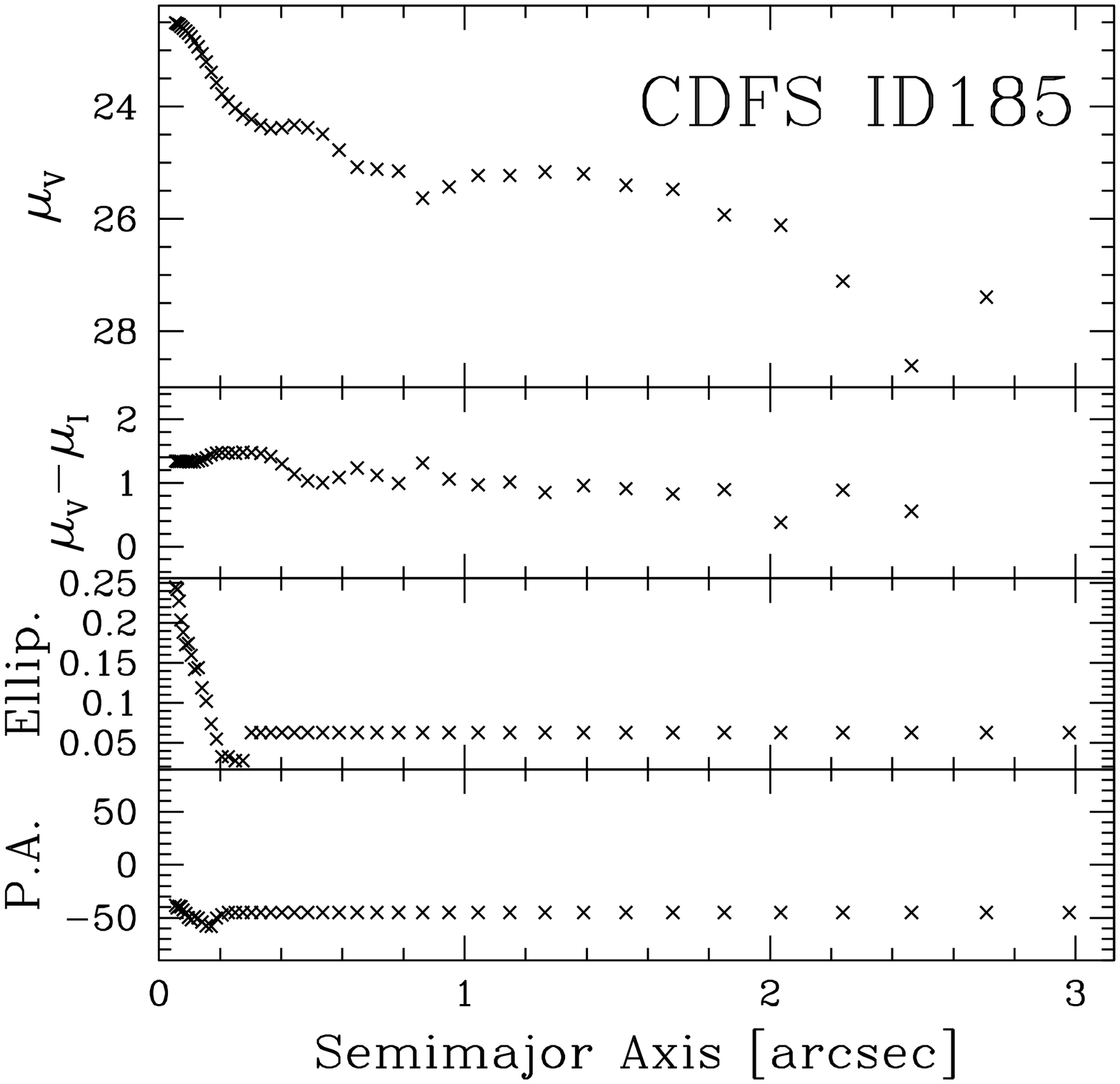,angle=0,width=\subplotwid}
\epsfig{figure=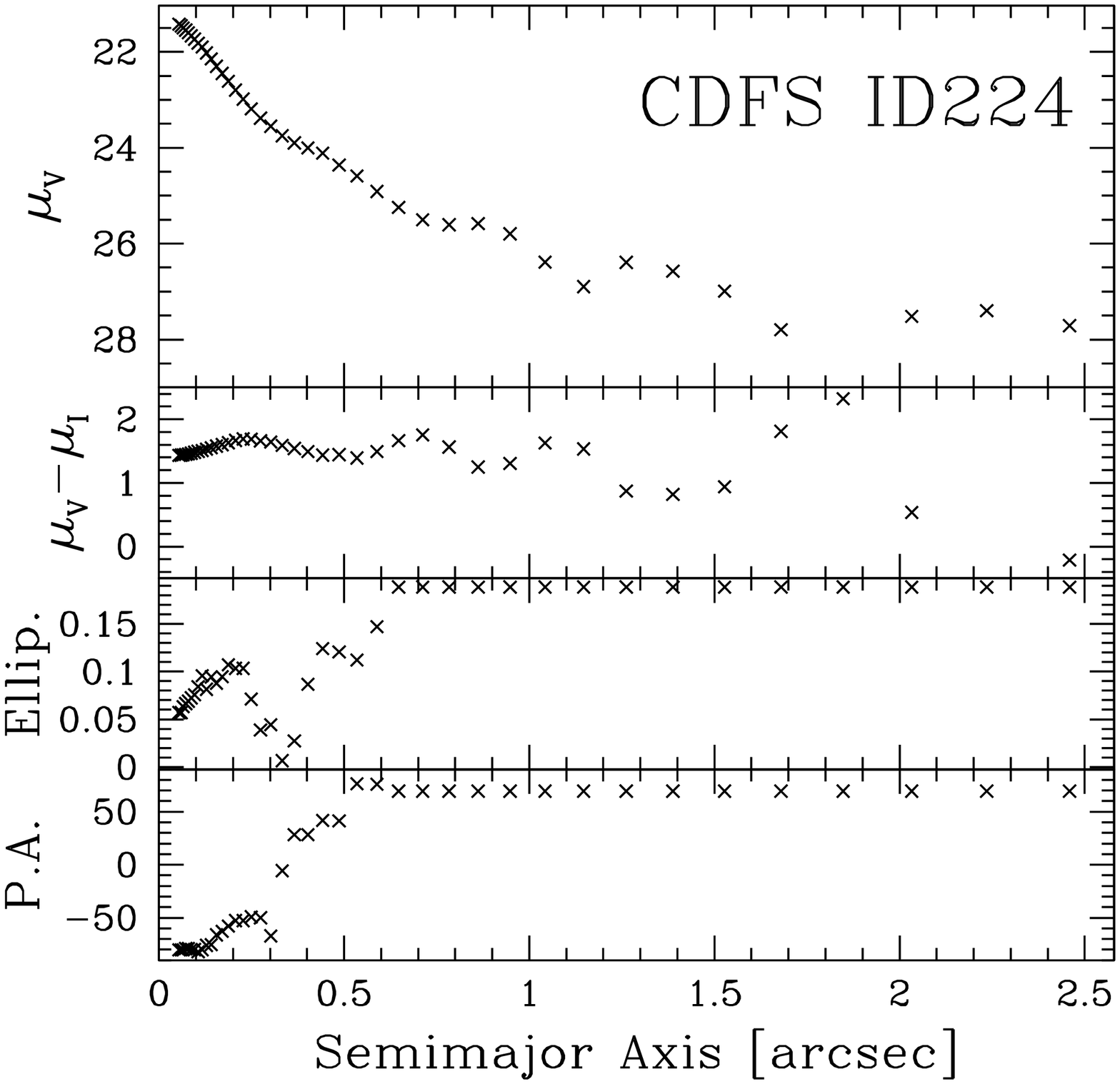,angle=0,width=\subplotwid}\\
\caption{Isophotal surface photometry of {\sl HST}-resolved $I\!<\!24$ counterparts to 
CDFS sources.
\label{sbproffig}}
\end{figure}
\clearpage
\begin{figure}[bpt]
\figurenum{\ref{sbproffig}}
\epsfig{figure=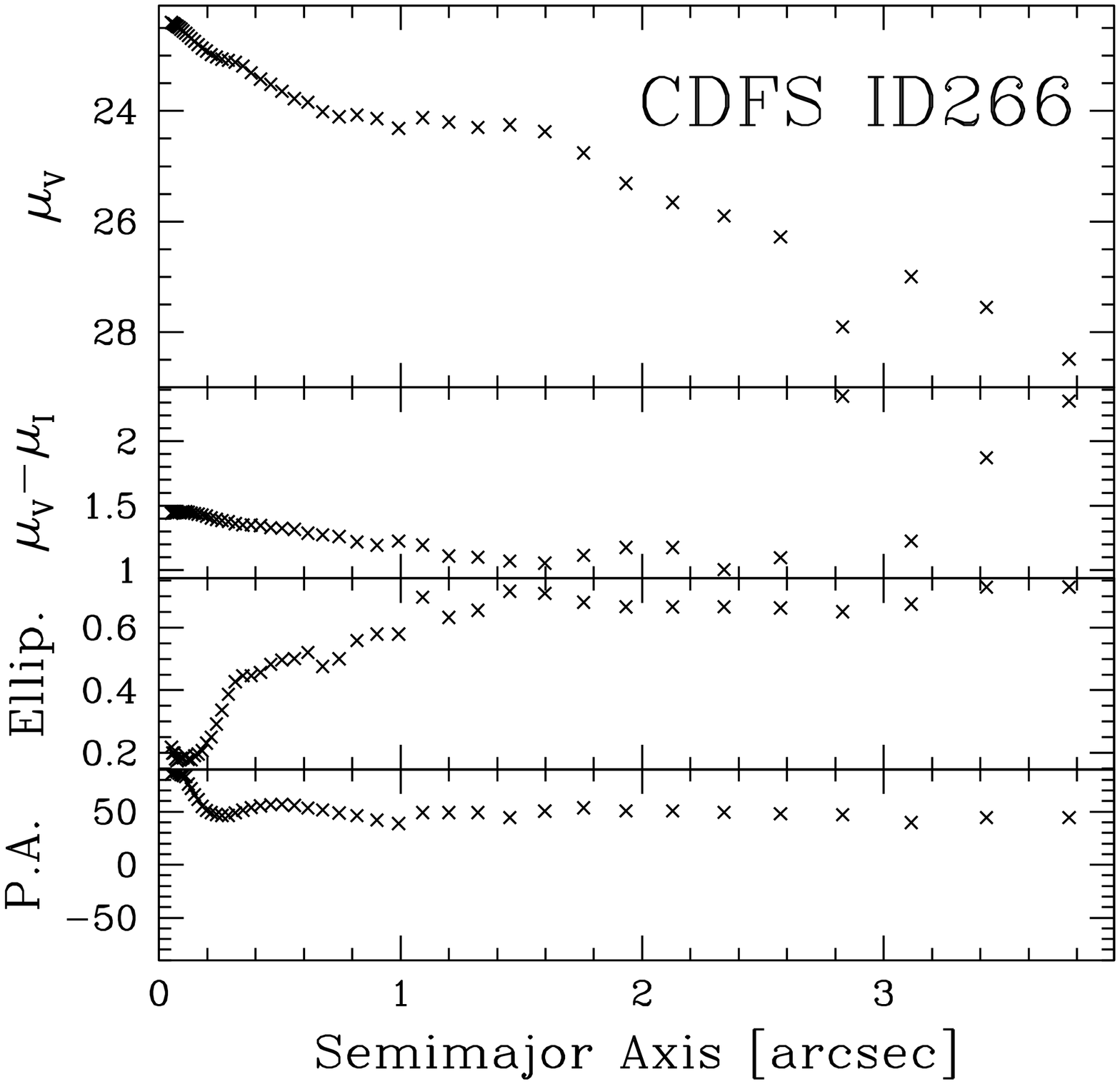,angle=0,width=\subplotwid}
\epsfig{figure=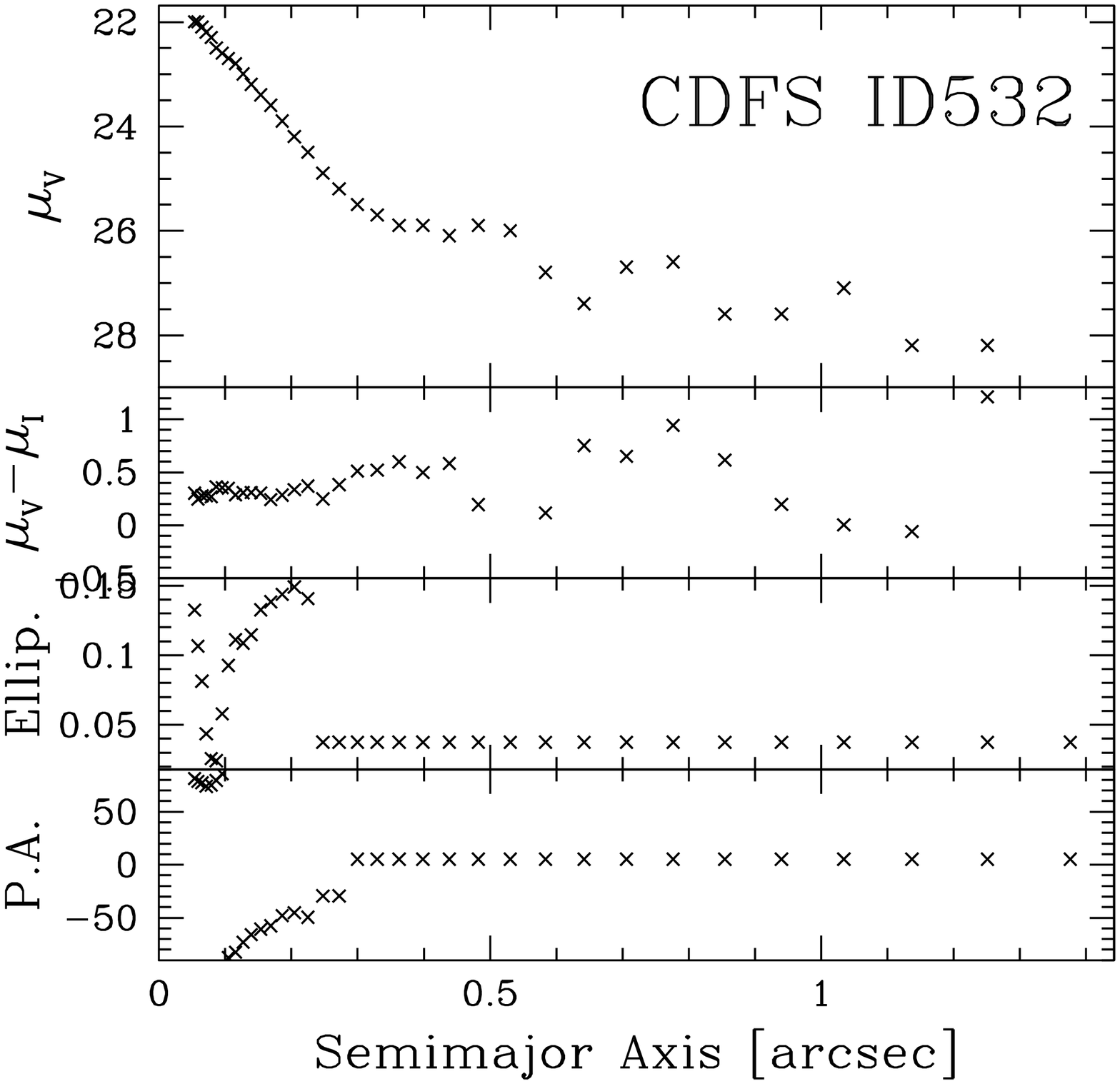,angle=0,width=\subplotwid}
\epsfig{figure=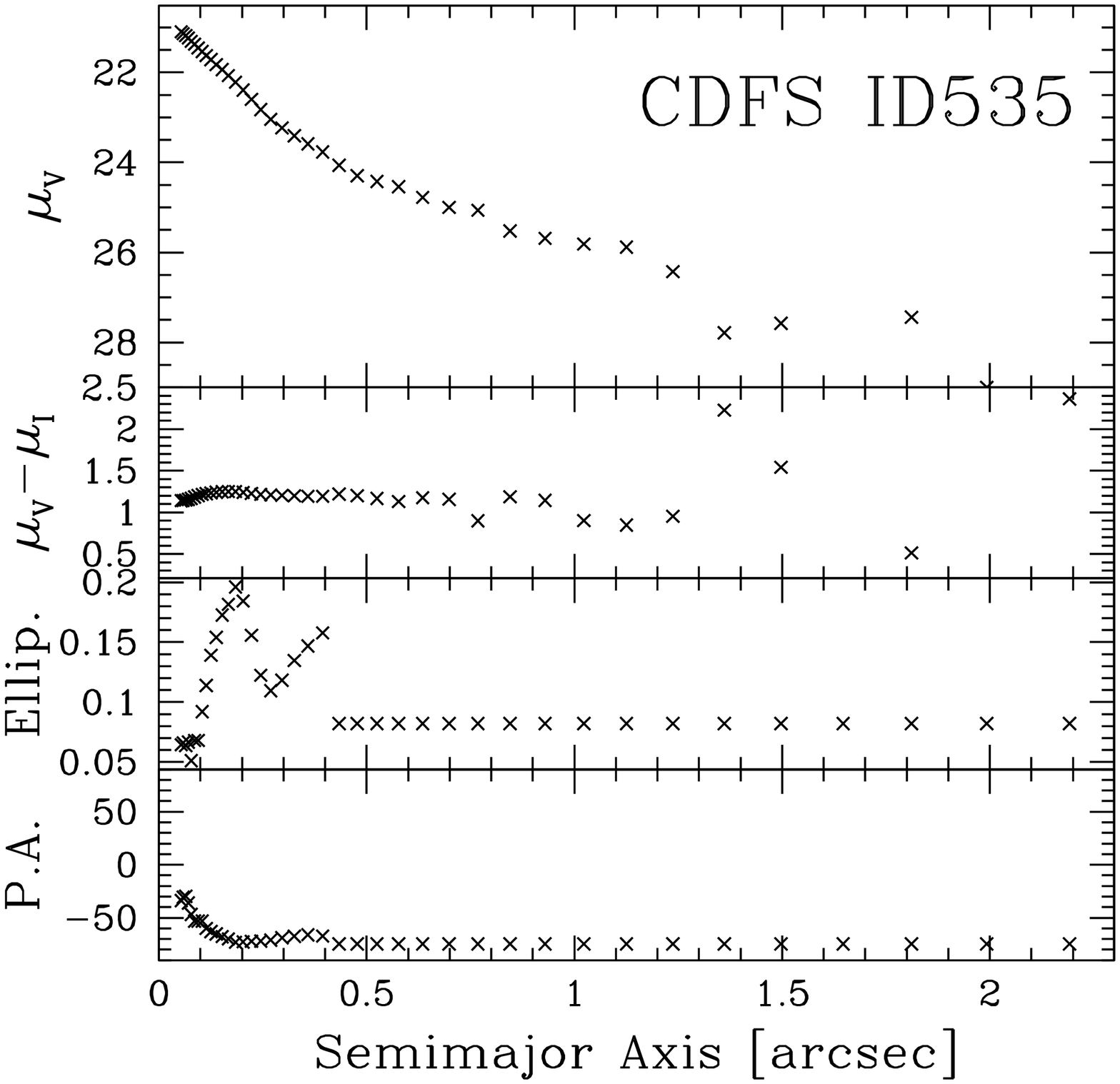,angle=0,width=\subplotwid}\\
\epsfig{figure=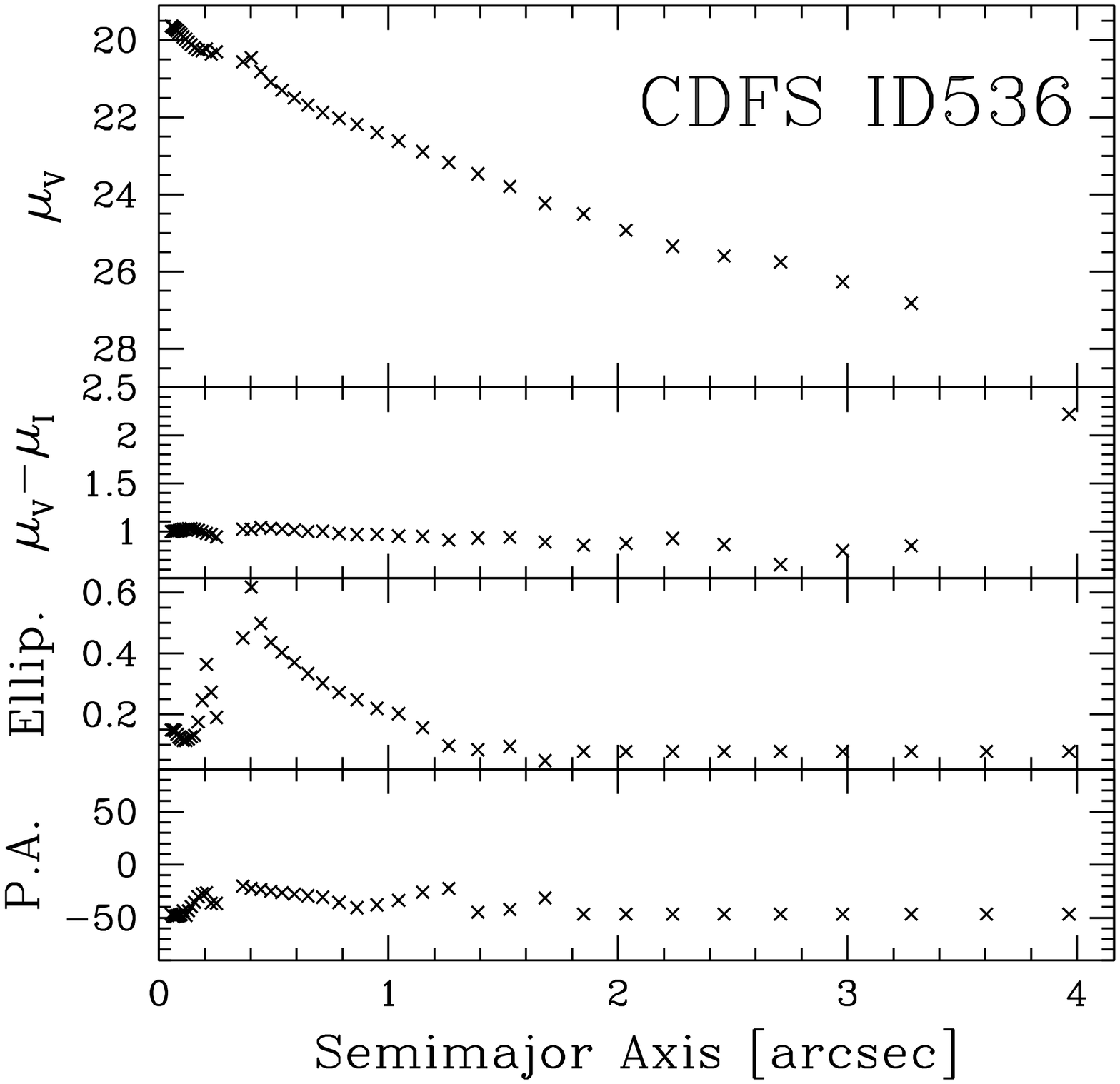,angle=0,width=\subplotwid}
\epsfig{figure=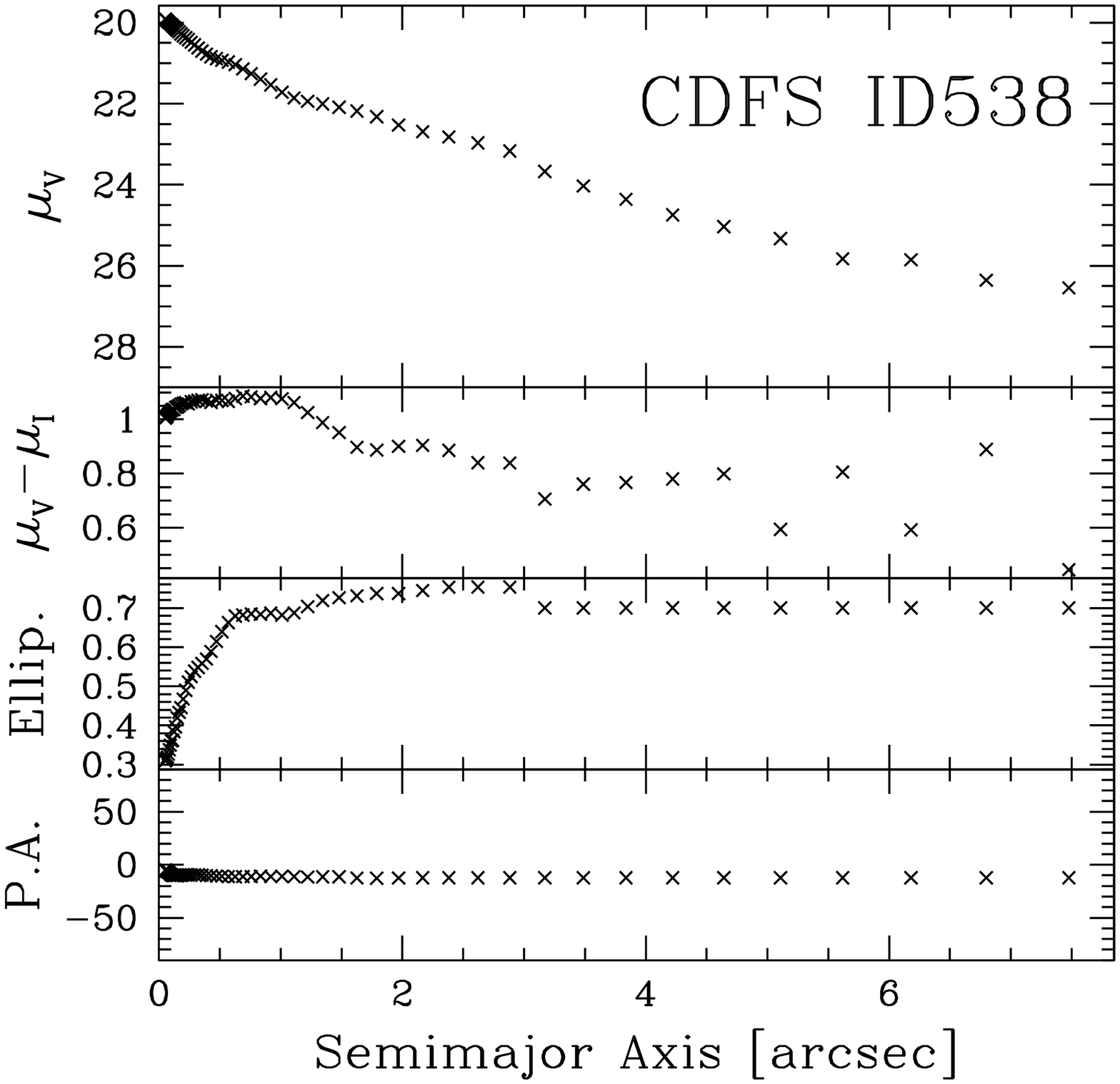,angle=0,width=\subplotwid}
\epsfig{figure=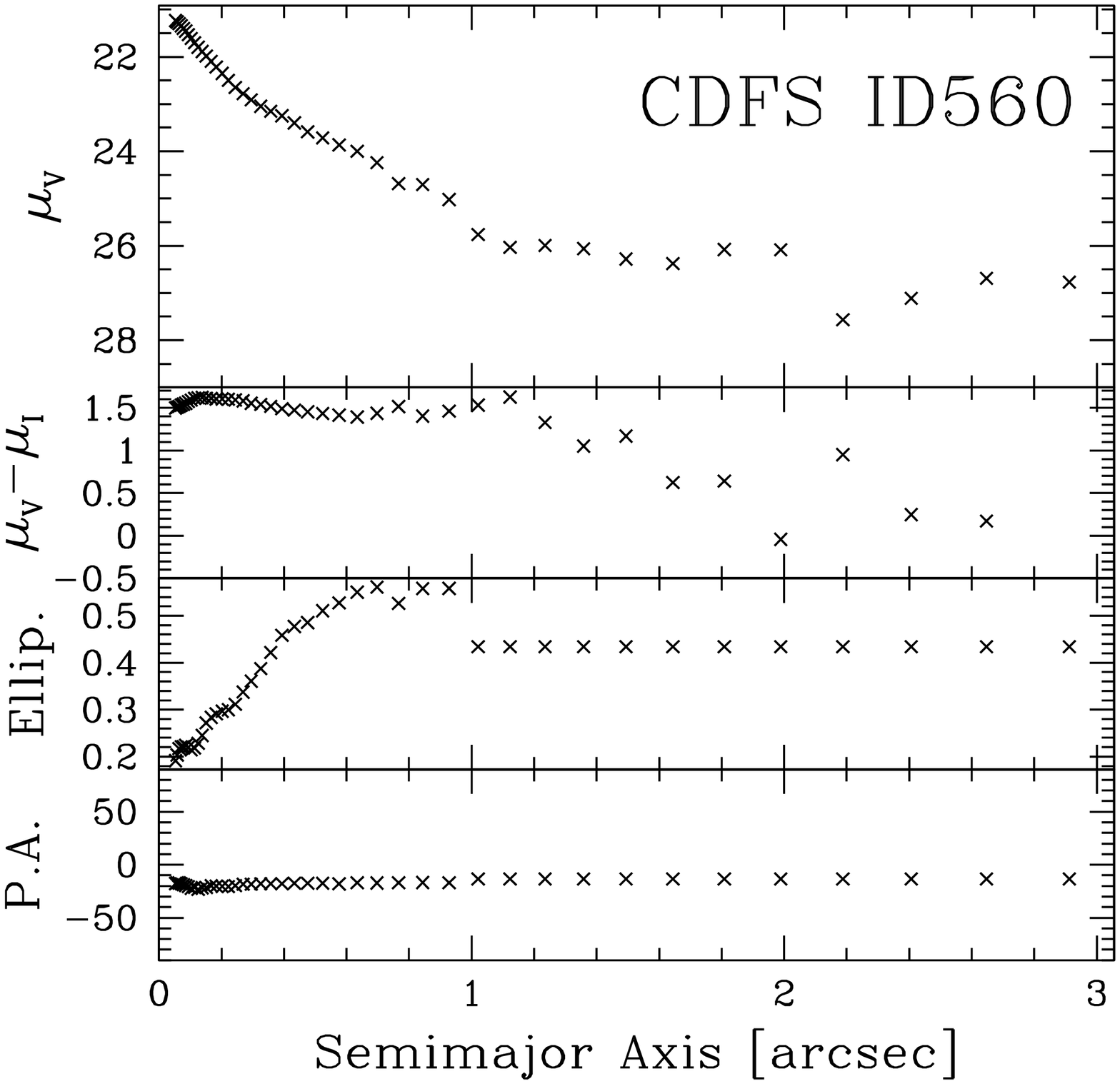,angle=0,width=\subplotwid}\\
\epsfig{figure=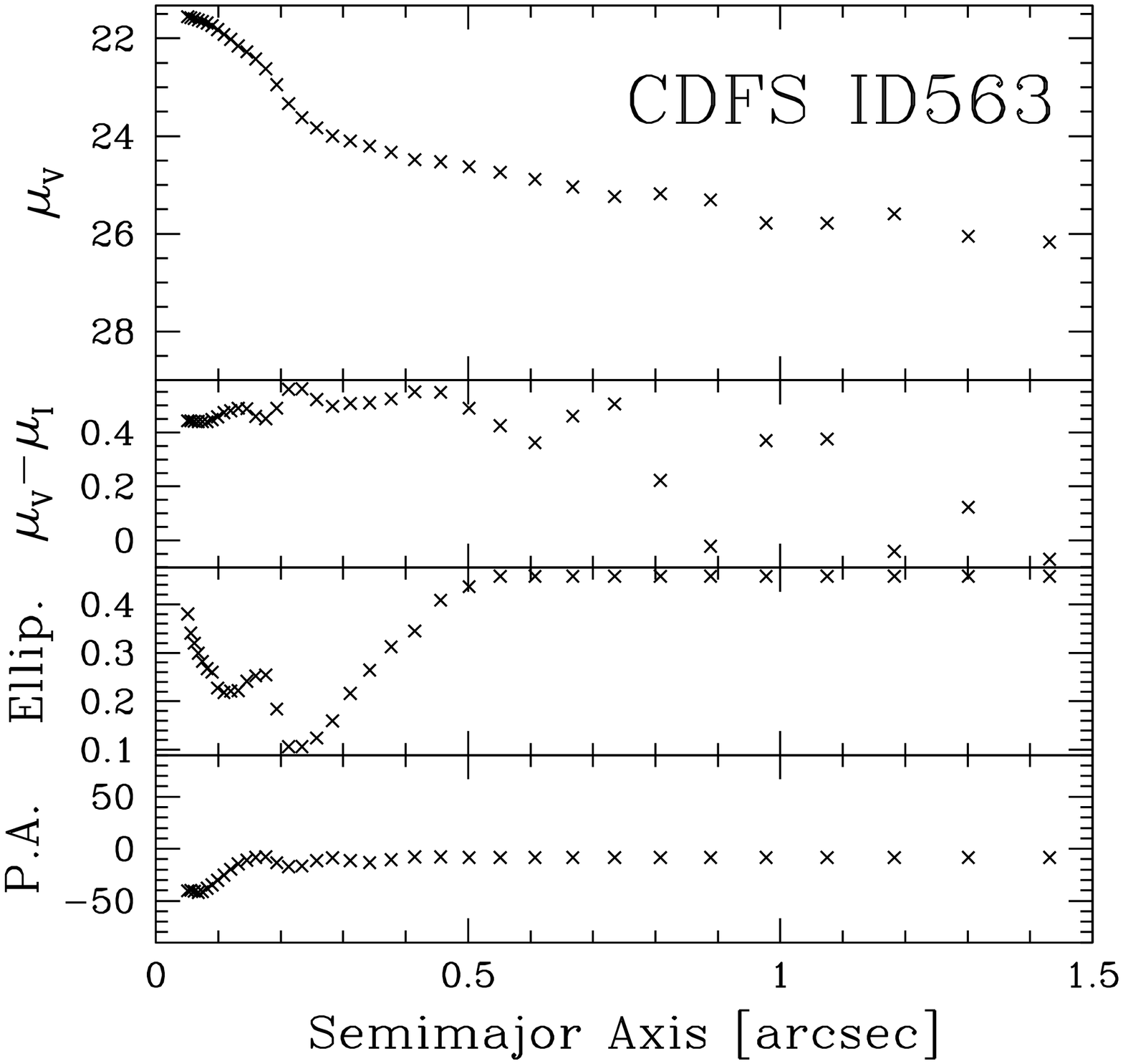,angle=0,width=\subplotwid}
\epsfig{figure=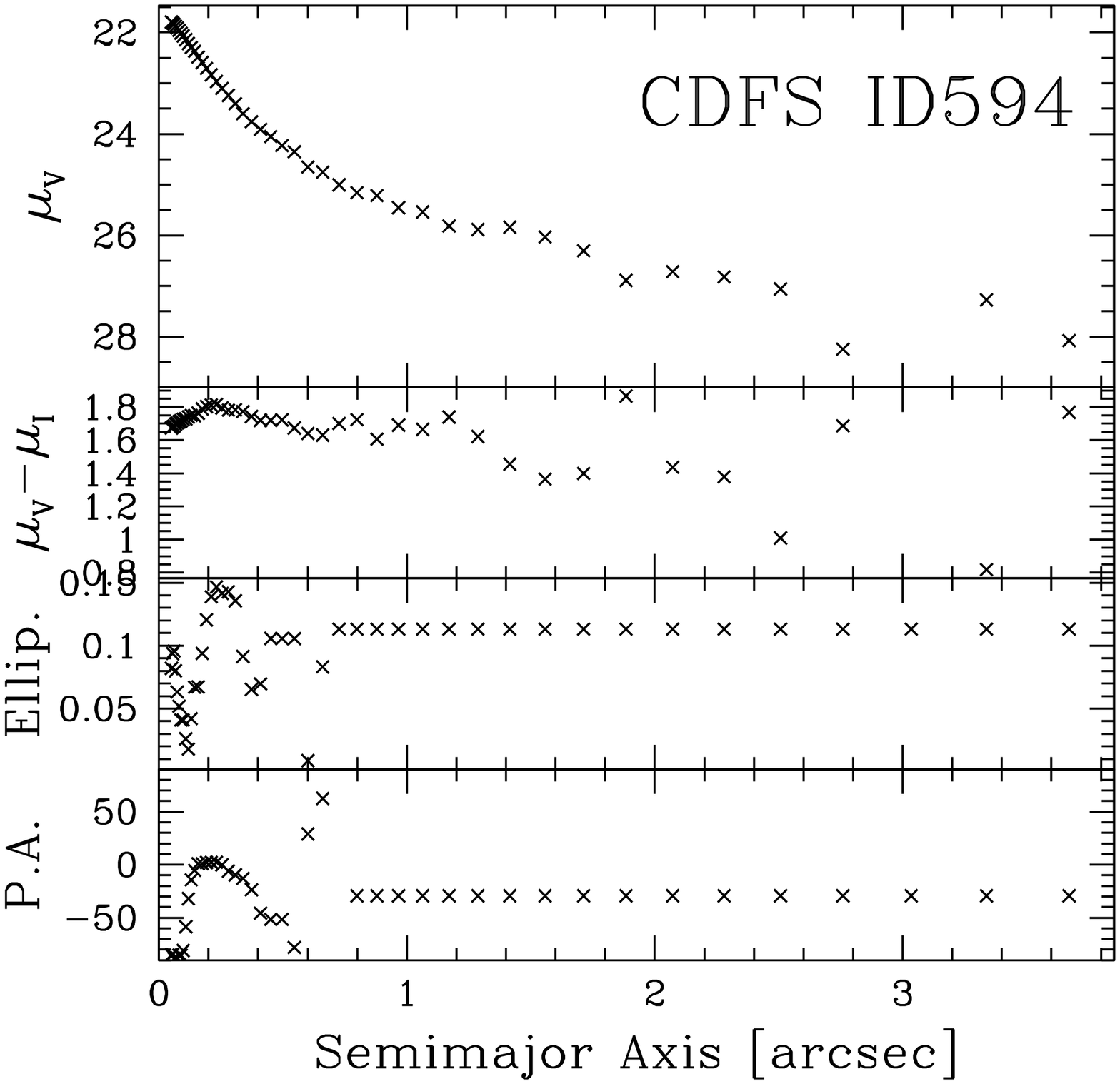,angle=0,width=\subplotwid}
\epsfig{figure=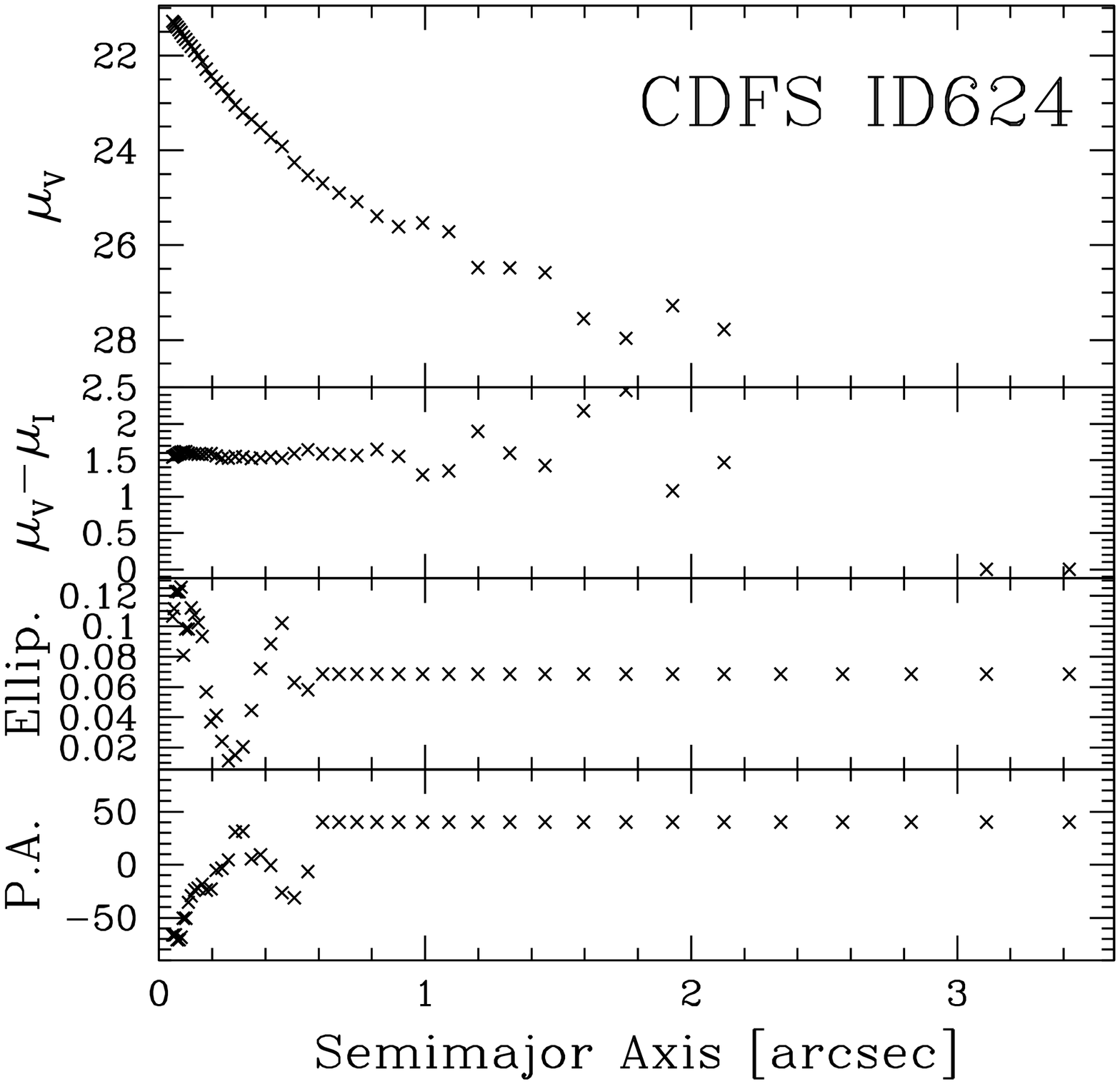,angle=0,width=\subplotwid}\\
\epsfig{figure=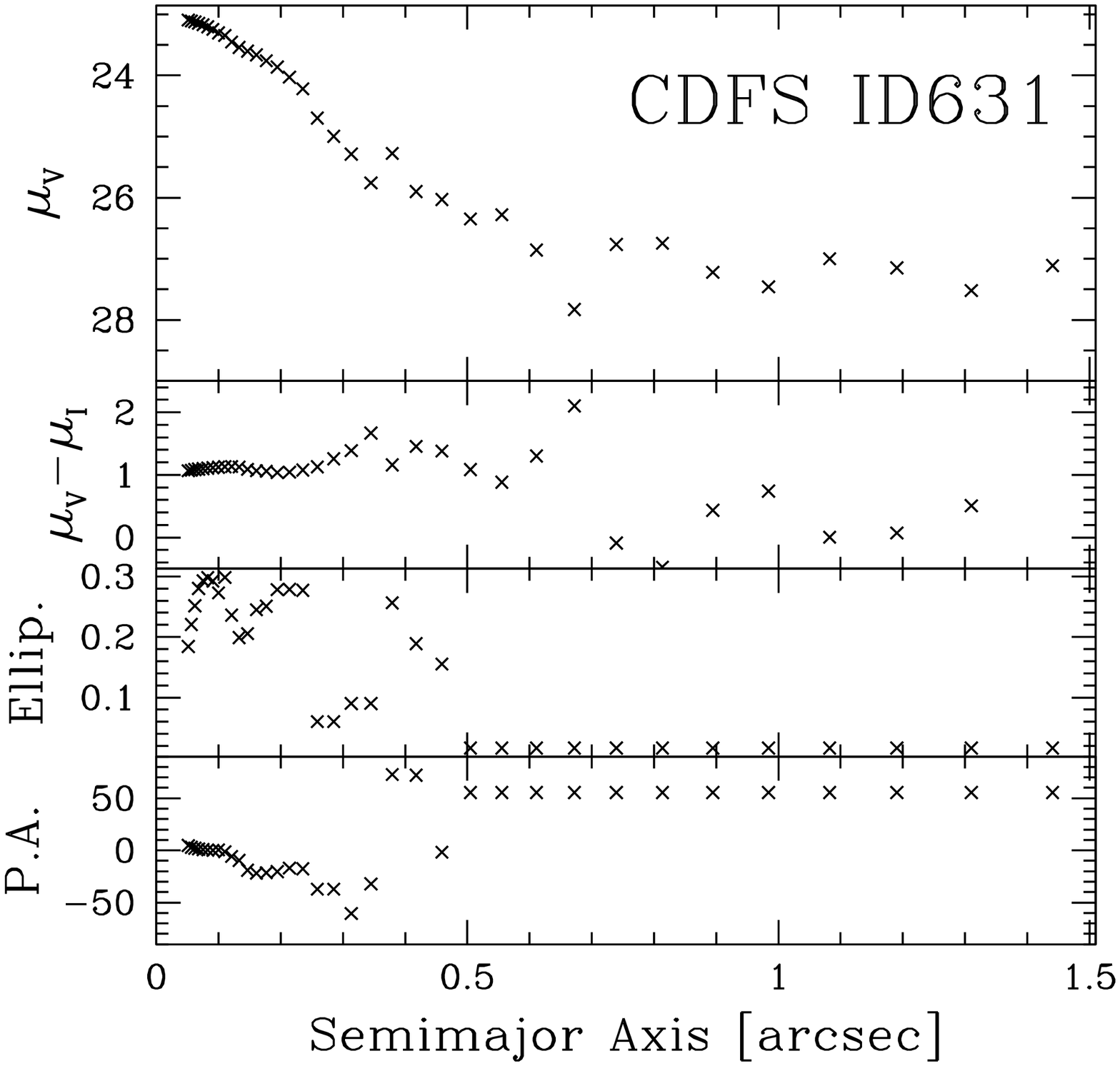,angle=0,width=\subplotwid}\\
\caption{Cont'd.}
\end{figure}

\begin{figure}[bpt]
\plotone{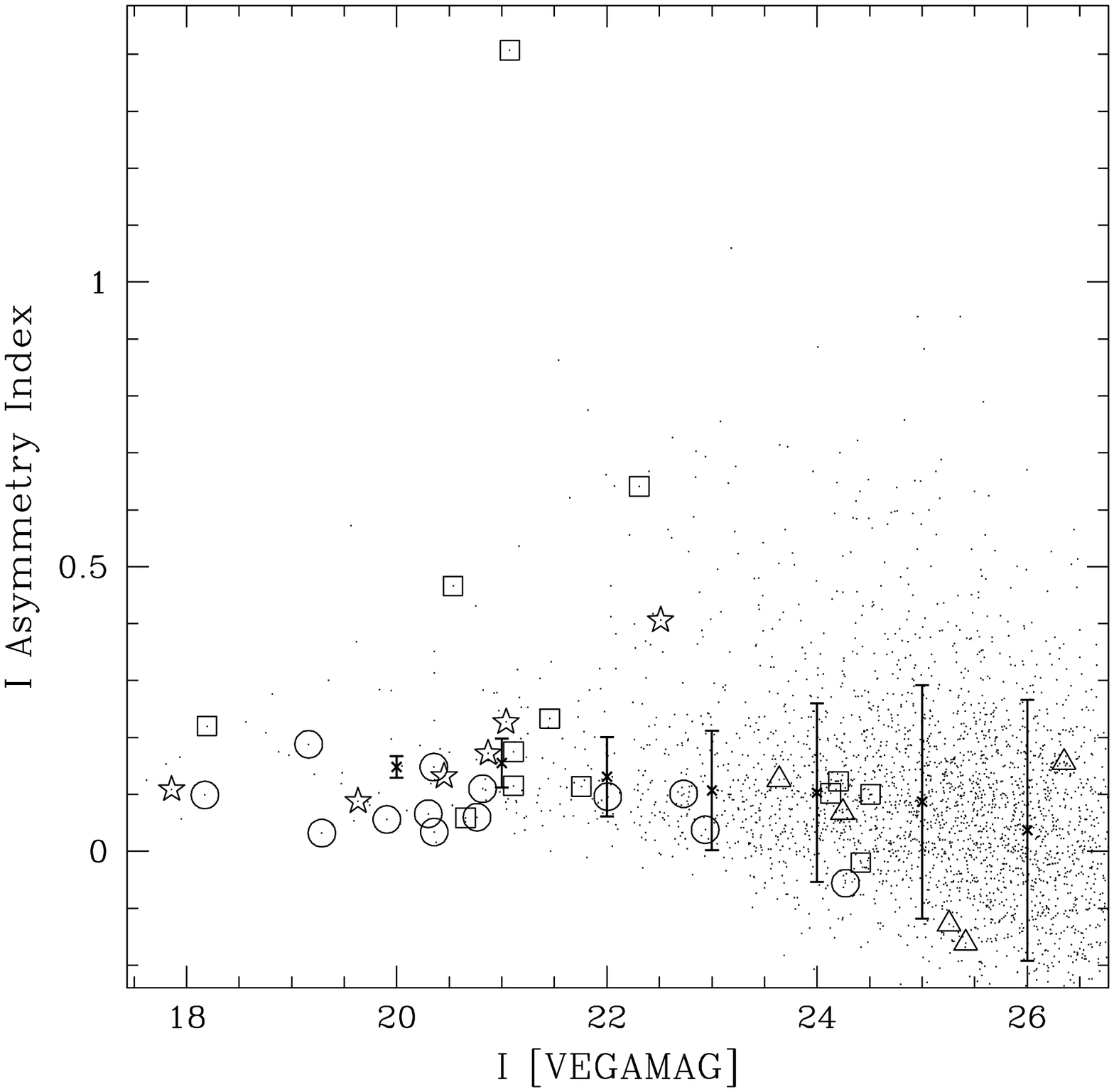}
\caption{Asymmetry index, $A$, of the $I$ surface brightness
  distributions for all detected sources in our three WFPC2 pointings
  (small dots) plotted as a function of total $I$ magnitude.  The
  sources detected by the 1~Ms \CXO\ observation are flagged by
  large symbols keyed to morphological type: E/S0 (circles); S/Irr
  (squares); unresolved (stars); and indeterminate (triangles).  The
  crosses and their associated error bars respectively denote the
  median $A$ and the median uncertainty in $A$ for field sources
  within successive 1.0-mag bins.\label{magasymfig}}
\end{figure}

\begin{figure}[bpt]
\plotone{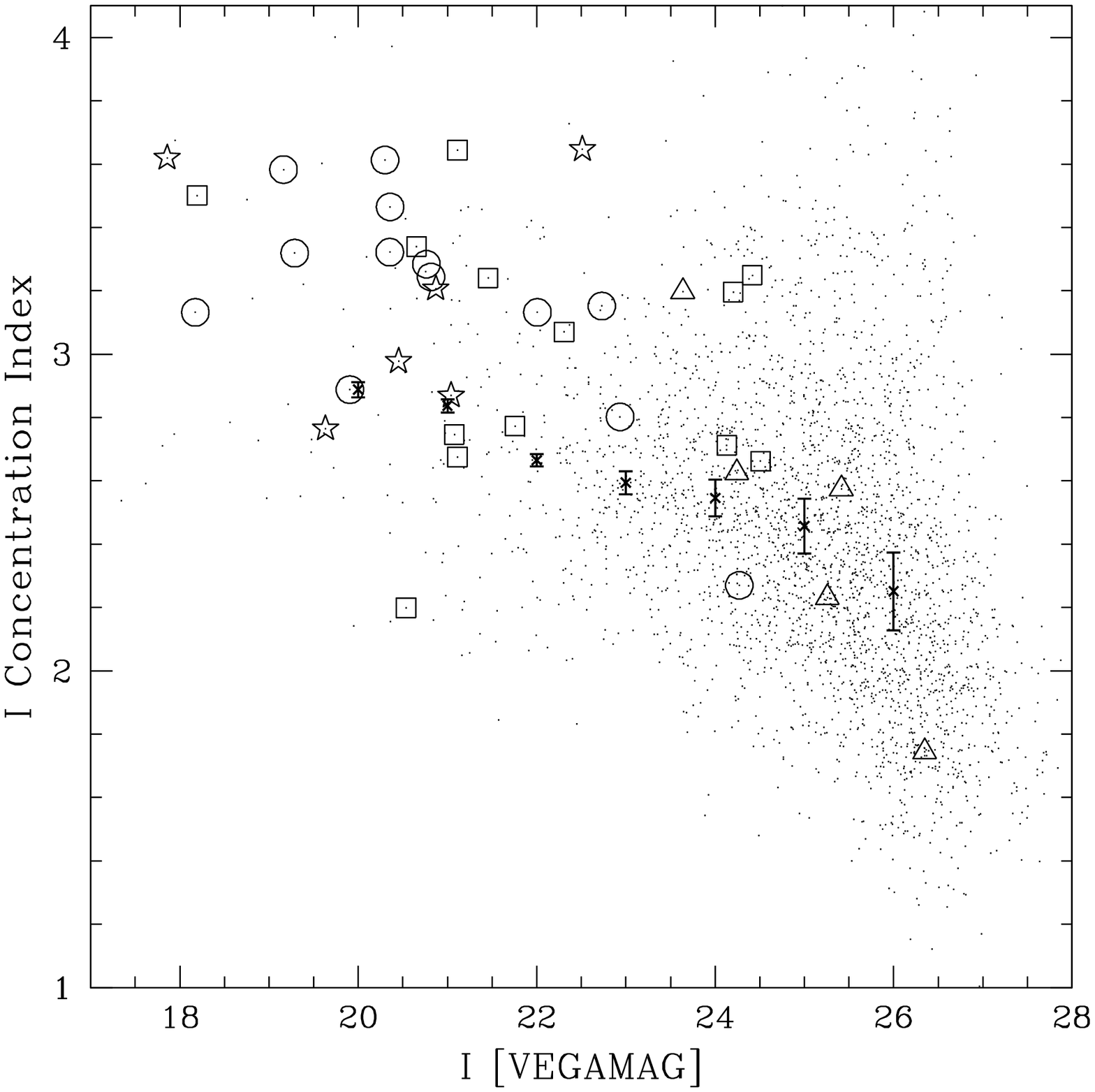}
\caption{Same as Fig.~\ref{magasymfig}, but for concentration 
  index $C$\label{concmagifig}}
\end{figure}

\begin{figure}[bpt]
\epsscale{.80}
\plotone{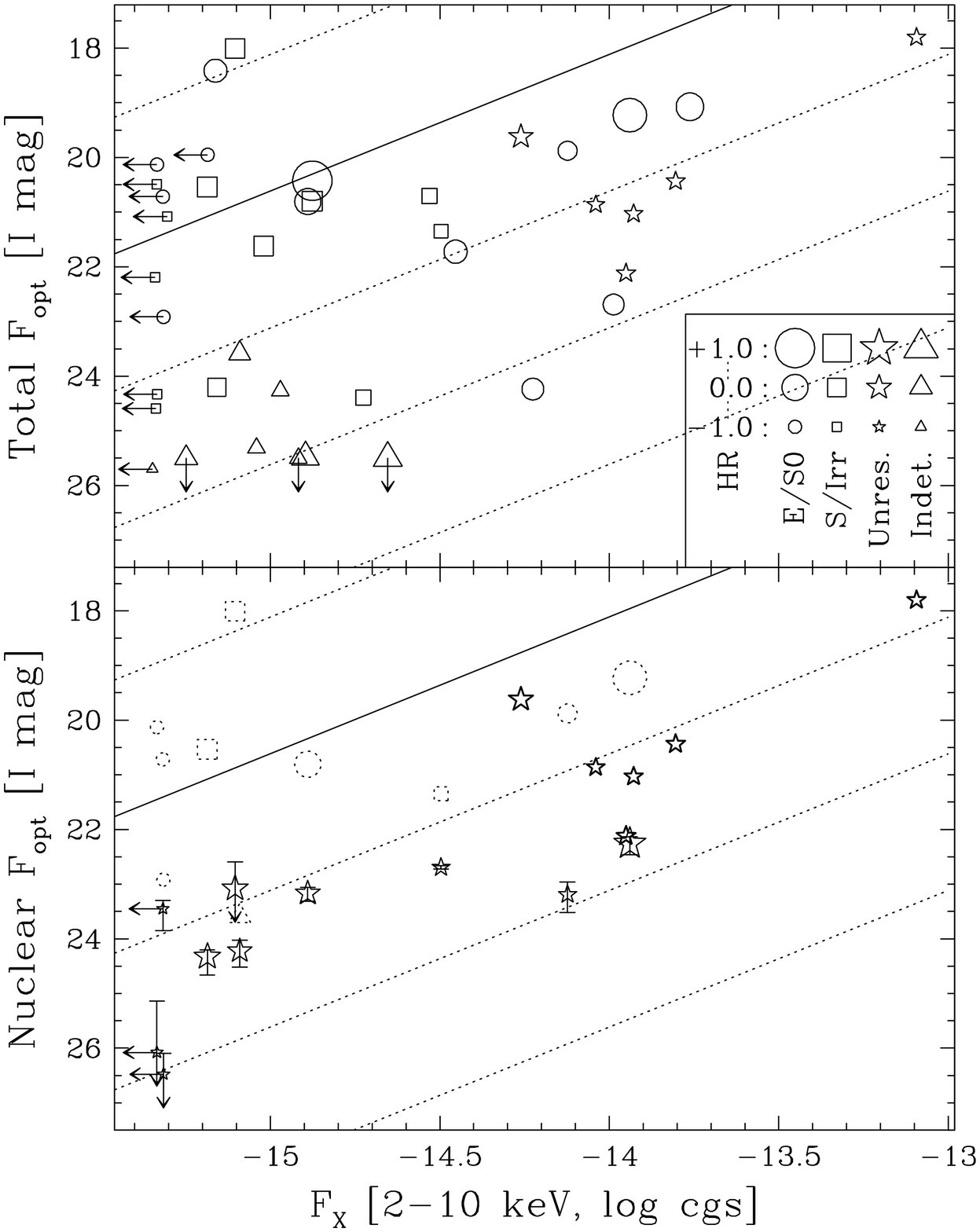}
\caption{$F_X$ versus~$F_{\rm opt}$ plots showing: a) integrated optical
flux for all \CXO\ sources.  X-ray hardness of the source is indicated
by symbol size, and host optical morphology is indicated by symbol
shape, as noted in the legend; b) optical unresolved nuclear flux for
the subset of $I<24$ counterparts best-fit with a finite nuclear
point-source.  Resolved hosts are now shown as open stars, below their
corresponding dotted symbol denoting total $F_{\rm opt}$ and host
morphology.  The unresolved hosts are now plotted as filled stars.
Diagonal lines are 1-dex intervals of $F_X/F_{\rm opt}$, with 
the solid line denoting $F_X/F_{\rm opt}=1$
\label{ptsrcfig}}
\end{figure}

\begin{figure}[bpt]
\epsscale{1.0}
\plotone{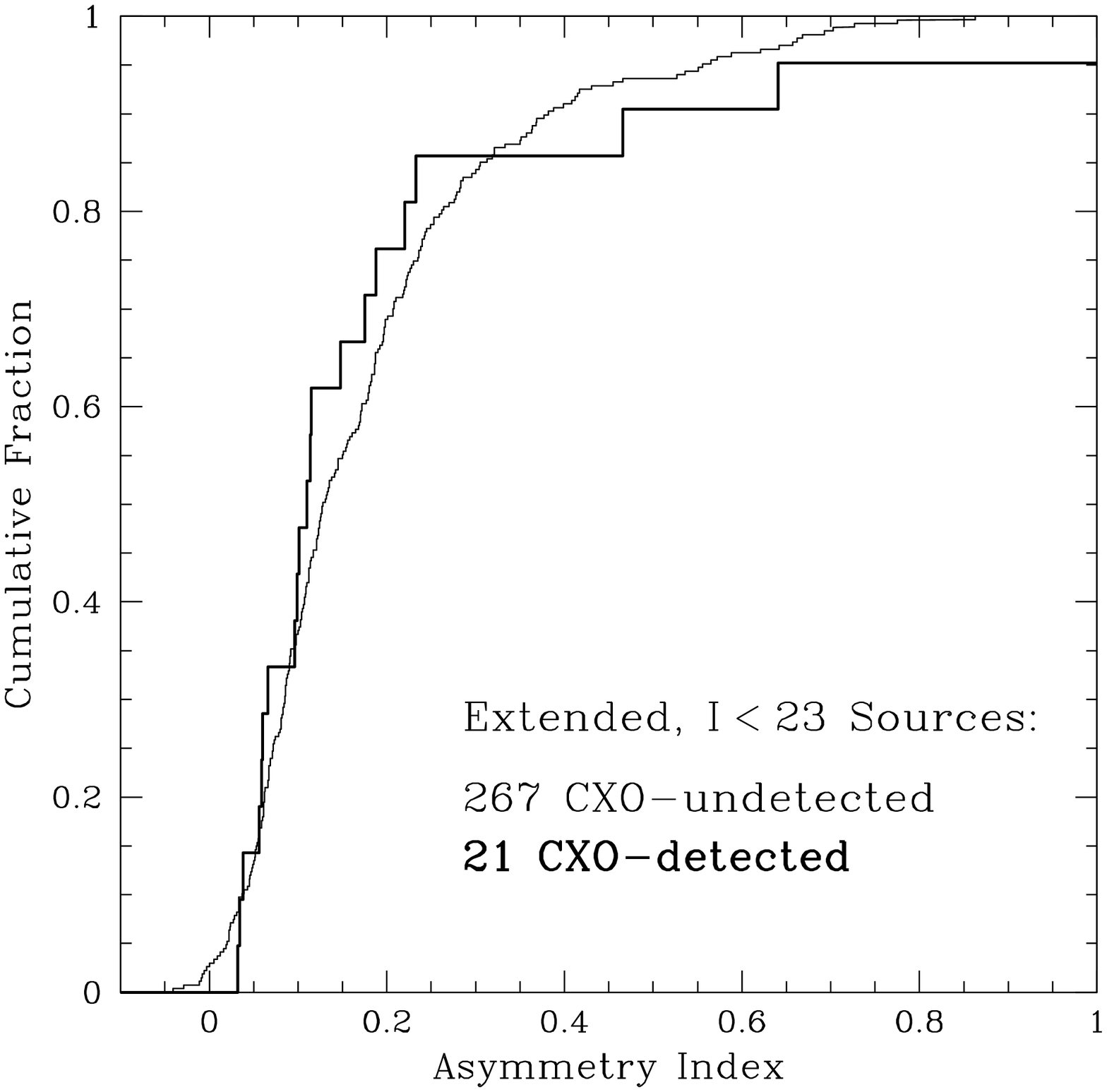}
\caption{Cumulative distribution functions of asymmetry index for bright
  resolved sources ($I<23$; $\eta_V<0.5$) in our three WFPC2 pointings in the
  Chandra Deep Field South.  The heavier line corresponds to those sources also
  detected in the 1~Ms \CXO\ exposure; the lighter line corresponds to
  the \CXO-undetected sources.\label{asymcdffig}}
\end{figure}

\begin{figure}[bpt]
\plotone{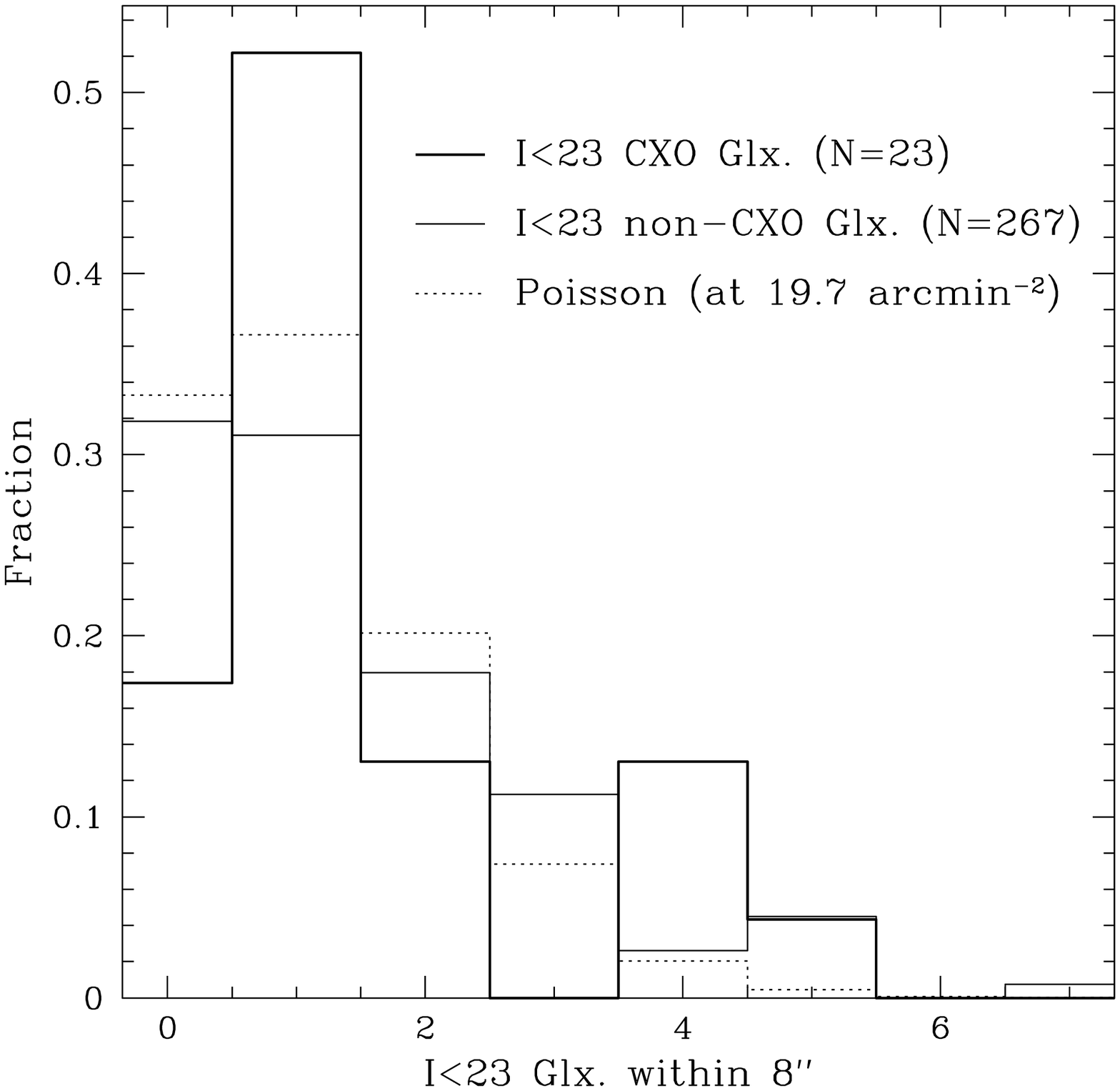}
\caption{Histogram of the number of nearby ($<\!8$\arcsec)
  $I<23$ galaxies appearing around the 21 resolved CDFS counterparts with 
$I<23$
  (heavy solid line) and around the 267 non-\CXO-detected galaxies in our
  catalog with $I<23$ (light solid line).  These two histograms are normalized
  by the sample sizes.  For comparison we also show the expected histogram if
  the $I<23$ galaxies were distributed at random given their observed number
  density (dotted line).\label{nnfig}}
\end{figure}

\begin{figure}[bpt]
\plotone{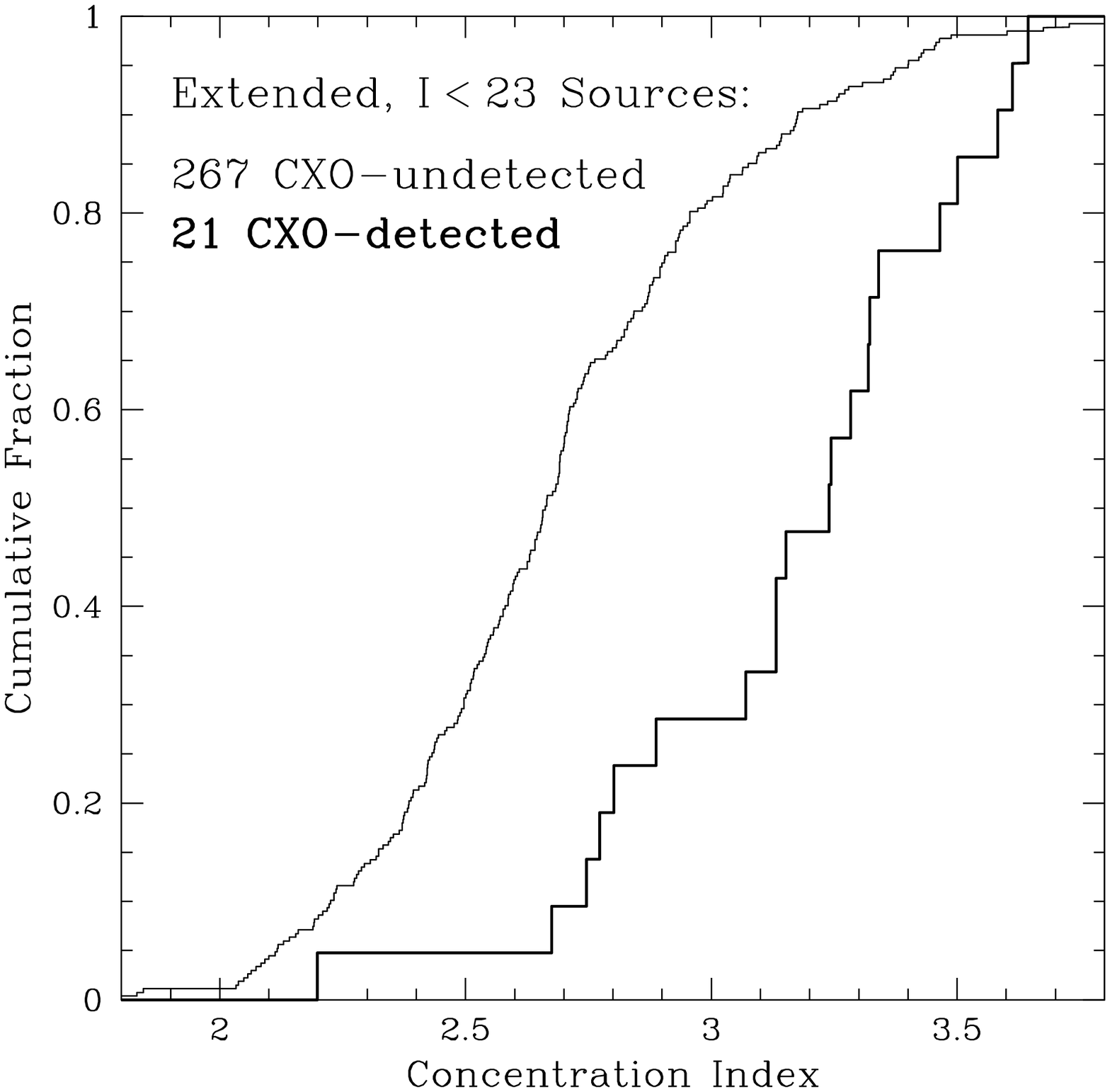}
\caption{Same as Fig.~\ref{asymcdffig}, but for concentration index $C$.\label{conccdffig}}
\end{figure}

\begin{figure}[bpt]
\plotone{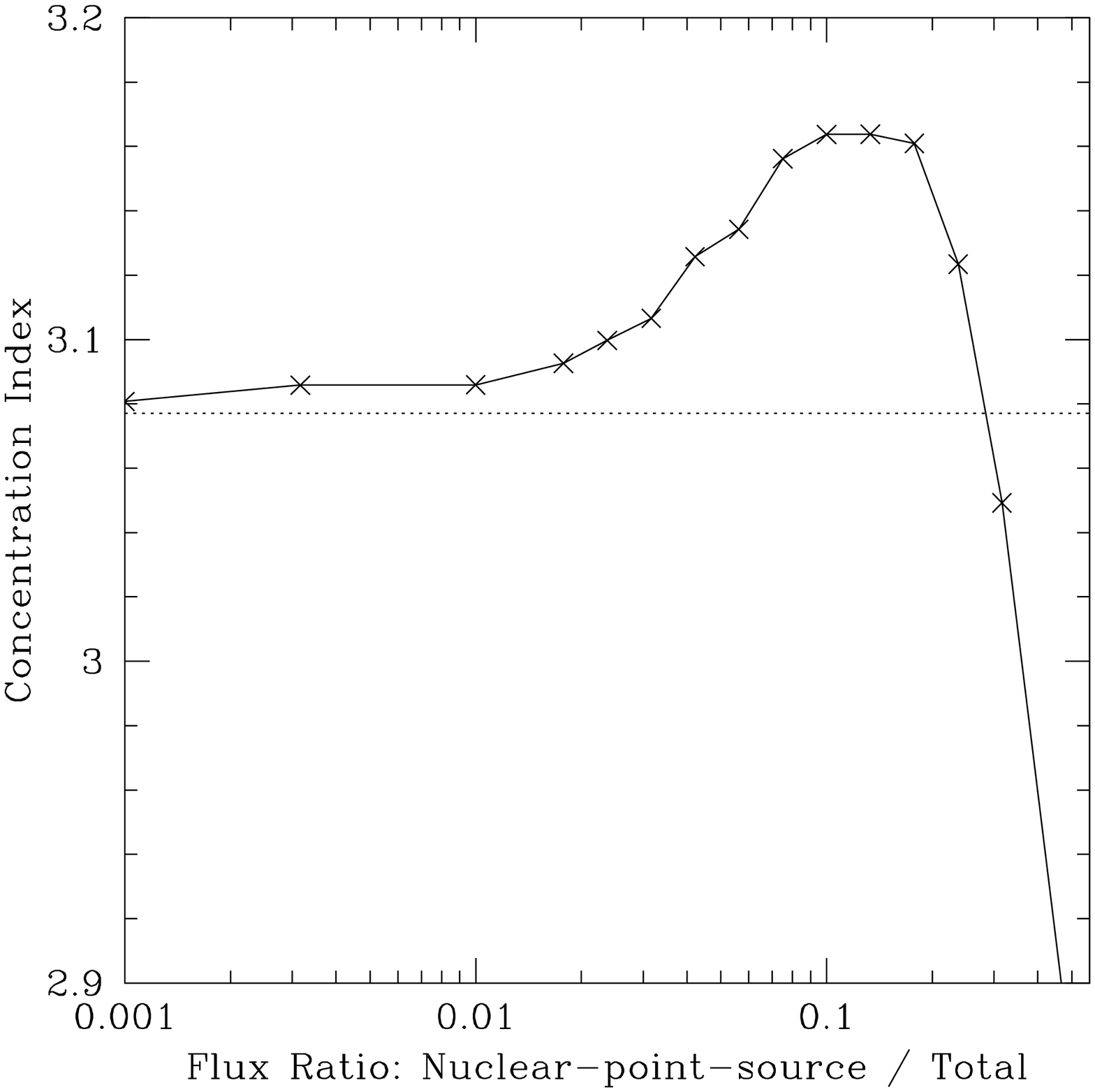}
\caption{Simulation of the concentration index ($C$) variation as a
function of nuclear point-source flux fraction.  The dotted line
indicates $C$ for the fiducial galaxy alone: an $I=22$ PSF-convolved
deVaucouleurs profile with $R_e=5$ pixels (0\farcs25).  The comparatively low $C$
of the PSF causes the turnover in the curve at large nuclear
point-source flux fractions.  The amplitude of the $C$ increase
remains small relative to the bias observed in
Figure~\ref{conccdffig}.
\label{concbiasplot}}
\end{figure}

\clearpage
\begin{deluxetable}{rrrrrrrrrr}
\tabletypesize{\scriptsize}
\tablecaption{\label{tab:hst_cat_pap3}
WFPC2 Source Catalog: Coordinates, Magnitudes, and Shape Parameters}
\tablewidth{0pt}
\tablehead{
\colhead{$\alpha_{\rm J2000}$} & \colhead{$\delta_{\rm J2000}$} & \colhead{F606W} & \colhead{F814W} & 
\colhead{$r_{0.5,I}$} & $\eta_V$ & $\eta_I$ & $A_I$ & $C_V$ & $C_I$ \\
\colhead{(deg)} & \colhead{(deg)} & \colhead{(mag)} & \colhead{(mag)} & \colhead{(arcsec)} &
&&&&
}
\startdata
53.007162 & $-27.769990$ & 27.53 & $>\!26.5$ &  0.16 & 0.02 & 0.63 & $ 0.06$ & 1.89 & 1.81 \\
53.007943 & $-27.771112$ & 26.43 & 25.16 &  0.24 & 0.00 & 0.00 & $ 0.03$ & 2.47 & 2.82 \\
53.008016 & $-27.770895$ & 27.34 & 26.07 &  0.20 & 0.80 & 0.01 & $-0.01$ & 3.34 & 2.88 \\
53.008110 & $-27.771046$ & 27.01 & 25.96 &  0.16 & 0.01 & 0.05 & $ 0.02$ & 2.13 & 2.43 \\
53.008349 & $-27.769775$ & 24.52 & 22.73 &  0.29 & 0.02 & 0.02 & $ 0.30$ & 2.28 & 2.63 \\
53.008414 & $-27.771456$ & 27.24 & $>\!26.5$ &  0.09 & 0.01 & 0.63 & $ 0.08$ & 2.56 & 1.83 \\
53.008675 & $-27.771753$ & 26.00 & 25.35 &  0.11 & 0.00 & 0.00 & $ 0.27$ & 2.60 & 2.20 \\
53.009305 & $-27.767180$ & 23.18 & 21.12 &  0.56 & 0.00 & 0.00 & $ 0.19$ & 2.25 & 2.39 \\
53.009428 & $-27.770188$ & 27.34 & $>\!26.5$ &  0.07 & 0.53 & 0.56 & $-0.25$ & 2.04 & 1.74 \\
53.009442 & $-27.763427$ & 24.83 & 23.08 &  0.38 & 0.00 & 0.02 & $ 0.76$ & 2.43 & 2.13 \\
53.009653 & $-27.767549$ & 26.48 & 25.89 &  0.13 & 0.00 & 0.00 & $-0.11$ & 3.02 & 2.76 \\
53.009702 & $-27.765063$ & 26.01 & 24.66 &  0.16 & 0.02 & 0.01 & $-0.05$ & 3.16 & 2.89 \\
53.009772 & $-27.766885$ & 27.15 & 25.55 &  0.12 & 0.00 & 0.00 & $-0.21$ & 2.89 & 2.16 \\
53.009885 & $-27.770993$ & 23.37 & 21.91 &  0.38 & 0.00 & 0.00 & $ 0.06$ & 2.46 & 2.60 \\
53.009959 & $-27.772112$ & 26.23 & 25.24 &  0.14 & 0.00 & 0.00 & $-0.09$ & 2.24 & 2.09 \\
53.010116 & $-27.764251$ & $>\!27.8$ & 26.05 &  0.10 & 0.06 & 0.01 & $-0.10$ & 1.67 & 2.29 \\
53.010253 & $-27.765836$ & 26.04 & 24.24 &  0.37 & 0.00 & 0.00 & $ 0.07$ & 3.85 & 3.02 \\
53.010534 & $-27.766503$ & 26.10 & 25.84 &  0.14 & 0.00 & 0.00 & $ 0.15$ & 2.12 & 1.86 \\
53.010582 & $-27.771649$ & 24.37 & 22.94 &  0.38 & 0.00 & 0.00 & $ 0.21$ & 2.04 & 2.42 \\
53.010584 & $-27.767012$ & 24.71 & 22.73 &  0.18 & 0.02 & 0.04 & $ 0.10$ & 3.12 & 3.15 \\
53.010590 & $-27.768191$ & 24.21 & 23.29 &  0.38 & 0.00 & 0.00 & $ 0.09$ & 2.33 & 2.23 \\
53.010664 & $-27.766375$ & 27.41 & $>\!26.5$ &  0.10 & 0.02 & 0.00 & $-0.18$ & 1.61 & 1.86 \\
53.010679 & $-27.769254$ & 27.25 & 25.65 &  0.15 & 0.00 & 0.00 & $-0.09$ & 2.09 & 2.86 \\
53.010681 & $-27.761041$ & $>\!27.8$ & 25.99 &  0.12 & 0.01 & 0.00 & $ 0.17$ & 1.62 & 1.79 \\
53.010846 & $-27.760243$ & 23.13 & 21.56 &  0.18 & 0.03 & 0.03 & $ 0.11$ & 2.69 & 2.82 \\
53.010918 & $-27.764613$ & 27.12 & 25.53 &  0.12 & 0.03 & 0.00 & $ 0.19$ & 2.07 & 2.77 \\
53.011001 & $-27.765044$ & 27.28 & 25.93 &  0.19 & 0.01 & 0.00 & $-0.14$ & 2.35 & 1.97 \\
53.011020 & $-27.766578$ & 26.36 & 24.87 &  0.20 & 0.00 & 0.00 & $ 0.11$ & 3.73 & 2.55 \\
53.011088 & $-27.769922$ & 27.15 & 25.91 &  0.17 & 0.00 & 0.01 & $-0.30$ & 3.68 & 2.87 \\
53.011158 & $-27.766516$ & 27.71 & 25.53 &  0.14 & 0.00 & 0.00 & $ 0.35$ & 3.14 & 1.91 \\
53.011179 & $-27.762714$ & 27.19 & $>\!26.5$ &  0.08 & 0.04 & 0.02 & $ 0.23$ & 1.92 & 2.48 \\
53.011219 & $-27.772550$ & $>\!27.8$ & 25.54 &  0.21 & 0.05 & 0.00 & $ 0.09$ & 2.48 & 1.91 \\
53.011291 & $-27.760761$ & 25.41 & 23.88 &  0.56 & 0.00 & 0.00 & $ 0.15$ & 2.69 & 2.08 \\
53.011320 & $-27.760891$ & 26.56 & 25.36 &  0.44 & 0.00 & 0.00 & $ 0.14$ & 2.56 & 3.10 \\
53.011347 & $-27.766480$ & $>\!27.8$ & $>\!26.5$ &  0.10 & 0.01 & 0.02 & $ 0.28$ & 1.86 & 2.25 \\
53.011536 & $-27.764139$ & 26.01 & 24.25 &  0.16 & 0.00 & 0.01 & $ 0.01$ & 2.36 & 2.44 \\
53.011569 & $-27.772637$ & 24.62 & 23.56 &  0.25 & 0.00 & 0.00 & $-0.03$ & 2.42 & 2.56 \\
53.011587 & $-27.758720$ & 27.52 & 26.39 &  0.15 & 0.00 & 0.01 & $-0.18$ & 2.22 & 1.78 \\
53.011632 & $-27.770297$ & 26.61 & 25.32 &  0.21 & 0.00 & 0.00 & $ 0.05$ & 3.27 & 2.63 \\
53.011753 & $-27.770756$ & 26.00 & 24.80 &  0.38 & 0.00 & 0.00 & $ 0.01$ & 3.24 & 2.21 \\
53.011806 & $-27.769385$ & 26.97 & 26.31 &  0.10 & 0.01 & 0.00 & $ 0.19$ & 2.04 & 1.86 \\
53.011876 & $-27.767014$ & 26.96 & $>\!26.5$ &  0.20 & 0.01 & 0.48 & $-0.06$ & 1.98 & 1.17 \\
\enddata
\tablecomments{
\baselineskip 12pt 
Table \ref{tab:hst_cat_pap3} is available in its entirety 
by request from the authors, and will be published in the electronic
edition of the {\it Astrophysical Journal}.
A portion is shown here for guidance regarding its form and content.}
\end{deluxetable}

\begin{deluxetable}{lccccccccccccl}
\rotate
\setlength{\tabcolsep}{0.02in}
\renewcommand{\arraystretch}{1.2}
\tabletypesize{\tiny}
\tablecaption{\label{tab:hst_cxo_pap3}%
WFPC2 CDFS Counterparts: X-ray and Optical Properties}
\tablehead{
\colhead{XID} & 
\colhead{IAU Designation} &
\colhead{$\log F_{2-10\,{\rm keV}}$\tablenotemark{a}} &
\colhead{$H.R.$\tablenotemark{b}} &
\colhead{F606W}& 
\colhead{F814W}&
\colhead{$r_{0.5,I}$} &
\colhead{$\eta_V$} &
\colhead{$\eta_I$} &
\colhead{$(f_{\rm nuc}/f_{\rm tot})_V$\tablenotemark{c}} &
\colhead{$(f_{\rm nuc}/f_{\rm tot})_I$\tablenotemark{c}} &
\colhead{$A_I$} &
\colhead{$C_I$} &
\colhead{Comments}\\
&
&
\colhead{(cgs)} &
&
\colhead{(mag)} &
\colhead{(mag)} &
\colhead{(\arcsec)} & 
&
&
&
&
&
&
}
\startdata
\phn36 & J033233.1$-$274548 & $-14.53$    & $-0.36$ & 22.95 & 21.08 & 0.71 & 0.00 & 0.00 & $0^{<0.0086}$ & $0^{<0.033}$ & \phs$1.407\!:\pm0.012$ & $2.75\pm0.01$ & In compact group\\
\phn38 & J033230.3$-$274505 & $-14.04$    & $-0.56$ & 22.18 & 20.87 & 0.16 & 0.69 & 0.03 & $\sim\!1$\tablenotemark{d} & $\sim\!1$\tablenotemark{d} & \phs$0.172\pm0.005$ & $3.21\pm0.04$ & \\
\phn39 & J033230.1$-$274530 & $-13.80$    & $-0.47$ & 21.34 & 20.45 & 0.14 & 0.87 & 0.15 & $\sim\!1$\tablenotemark{d} & $\sim\!1$\tablenotemark{d} & \phs$0.132\pm0.003$ & $2.98\pm0.04$ & \\
\phn52 & J033217.2$-$274304 & $-14.12$    & $-0.54$ & 21.57 & 19.91 & 0.25 & 0.03 & 0.03 & $0.10^{<0.22}$ & $0.049^{+0.012}_{-0.012}$ & \phs$0.056\pm0.005$ & $2.89\pm0.03$ & \\
\phn56 & J033213.3$-$274241 & $-13.76$    & $+0.11$ & 20.83 & 19.16 & 0.35 & 0.03 & 0.03 & $0.0096^{<0.066}$ & $0^{<0.011}$ & \phs$0.188\pm0.004$ & $3.58\pm0.03$ & \\
\phn58 & J033211.8$-$274629 & $-14.73$    & $-0.34$ & 25.95 & 24.42 & 0.19 & 0.00 & 0.02 & \nodata & \nodata & $-0.020\pm0.162$ & $3.25\pm0.18$ & \\
\phn60 & J033211.0$-$274415 & $-13.93$    & $-0.48$ & 22.39 & 21.04 & 0.13 & 0.54 & 0.41 & $\sim\!1$\tablenotemark{d} & $\sim\!1$\tablenotemark{d} & \phs$0.227\pm0.004$ & $2.87\pm0.04$ & \\
\phn61 & J033210.6$-$274309 & $-13.95$    & $-0.44$ & 25.09 & 22.51 & 0.20 & 0.83 & 0.73 & $\sim\!1$\tablenotemark{d} & $\sim\!1$\tablenotemark{d} & \phs$0.406\pm0.070$ & $3.65\pm0.04$ & \\
\phn62 & J033209.5$-$274807 & $-14.26$    & $-0.04$ & 20.87 & 19.63 & 0.12 & 0.91 & 0.89 & $\sim\!1$\tablenotemark{d} & $\sim\!1$\tablenotemark{d} & \phs$0.088\pm0.001$ & $2.77\pm0.03$ & \\
\phn63 & J033208.7$-$274735 & $-13.09$    & $-0.49$ & 19.21 & 17.86 & 0.15 & 0.93 & 0.88 & $\sim\!1$\tablenotemark{d} & $\sim\!1$\tablenotemark{d} & \phs$0.109\pm0.000$ & $3.62\pm0.05$ & \\
\phn64 & J033208.1$-$274658 & $-14.23$    & $-0.31$ & 25.32 & 24.27 & 0.14 & 0.02 & 0.01 & \nodata & \nodata & $-0.056\pm0.090$ & $2.27\pm0.04$ & \\
\phn66 & J033203.7$-$274604 & $-13.94$    & $+0.56$ & 21.40 & 19.29 & 0.37 & 0.03 & 0.03 & $0^{<0.016}$ & $0^{<0.011}$ & \phs$0.032\pm0.006$ & $3.32\pm0.03$ & \\
\phn67 & J033202.5$-$274601 & $-13.99$    & $-0.40$ & 24.71 & 22.73 & 0.18 & 0.02 & 0.04 & $0.35^{<0.49}$ & $0.46^{+0.018}_{-0.025}$ & \phs$0.101\pm0.037$ & $3.15\pm0.03$ & \\
\phn78 & J033230.1$-$274524 & $-14.50$    & $-0.54$ & 23.04 & 21.45 & 0.26 & 0.37 & 0.03 & $0.655^{+0.027}_{-0.20}$ & $0.357^{+0.008}_{-0.009}$ & \phs$0.233\pm0.021$ & $3.24\pm0.02$ & \\
\phn81 & J033226.0$-$274515 & $-15.04$    & $-0.43$ & 26.41 & 25.26 & 0.10 & 0.13 & 0.13 & \nodata & \nodata & $-0.128\pm0.115$ & $2.23\pm0.08$ & \\
\phn83 & J033215.0$-$274225 & $-14.45$    & $-0.22$ & 23.80 & 22.01 & 0.25 & 0.03 & 0.03 & $0^{<0.0499}$ & $0.0378^{<0.0544}$ & \phs$0.096\pm0.032$ & $3.13\pm0.03$ & \\
\phn86 & J033233.9$-$274521 & $-15.16$    & $-0.04$ & 25.32 & 24.13 & 0.31 & 0.00 & 0.00 & \nodata & \nodata & \phs$0.102\pm0.202$ & $2.71\pm0.05$ & \\
\phn89 & J033208.3$-$274153 & $-14.97$    & $-0.44$ & 25.23 & 24.24 & 0.12 & 0.10 & 0.04 & \nodata & \nodata & \phs$0.068\pm0.068$ & $2.63\pm0.07$ & \\
149    & J033212.3$-$274621 & $-15.02$    & $+0.12$ & 23.85 & 21.76 & 0.32 & 0.03 & 0.03 & $0^{<0.064}$ & $0^{<0.088}$ & \phs$0.114\pm0.043$ & $2.77\pm0.02$ & \\
155    & J033208.0$-$274240 & $-14.88$    & $+0.17$ & 23.06 & 21.11 & 0.47 & 0.00 & 0.03 & $0^{<0.0053}$ & $0^{<0.031}$ & \phs$0.175\pm0.046$ & $3.64\pm0.04$ & \\
173    & J033216.8$-$274327 & $<\!-15.30$ & $-1.00$ & 22.88 & 21.11 & 0.22 & 0.03 & 0.03 & $0^{<0.026}$ & $0^{<0.043}$ & \phs$0.115\pm0.011$ & $2.67\pm0.02$ & \\
185    & J033211.0$-$274343 & $-15.19$    & $+0.14$ & 22.39 & 20.54 & 0.62 & 0.00 & 0.03 & $0.026^{+0.0018}_{-0.0018}$ & $0.024^{+0.0067}_{-0.0037}$ & \phs$0.466\!:\pm0.056$ & $2.20\pm0.02$ & Truncated by chip edge\\
224    & J033228.8$-$274621 & $<\!-15.32$ & $-1.00$ & 22.97 & 20.76 & 0.27 & 0.03 & 0.03 & $0^{<0.018}$ & $0.084^{+0.013}_{-0.026}$ & \phs$0.060\pm0.014$ & $3.28\pm0.03$ & \\
266    & J033214.0$-$274249 & $-14.88$    & $+1.00$ & 22.29 & 20.35 & 0.69 & 0.00 & 0.03 & $0^{<0.0036}$ & $0^{<0.0036}$ & \phs$0.148\pm0.045$ & $3.32\pm0.01$ & \\
515    & J033232.2$-$274652 & $-14.90$    & $+0.42$ & 27.78 & 25.41 & 0.19 & 0.01 & 0.00 & \nodata & \nodata & $-0.161\pm0.253$ & $2.57\pm0.10$ & \\
532    & J033214.2$-$274231 & $-15.09$    & $-0.05$ & 24.81 & 23.64 & 0.13 & 0.22 & 0.04 & $0.93^{+0.0056}_{-0.18}$ & $0.58^{+0.12}_{-0.14}$ & \phs$0.126\pm0.042$ & $3.20\pm0.07$ & \\
535    & J033211.5$-$274650 & $-14.89$    & $+0.02$ & 22.69 & 20.82 & 0.31 & 0.03 & 0.03 & $0.013^{<0.11}$ & $0.12^{+0.012}_{-0.015}$ & \phs$0.110\pm0.017$ & $3.24\pm0.03$ & \\
536    & J033210.9$-$274235 & $-15.16$    & $-0.24$ & 19.97 & 18.17 & 0.59 & 0.03 & 0.03 & $0^{<0.0076}$ & $0^{<0.0086}$ & \phs$0.099\pm0.005$ & $3.13\pm0.02$ & Double nucleus, 0\farcs3 sep. \\
538    & J033208.6$-$274649 & $-15.10$    & $+0.12$ & 19.90 & 18.19 & 0.85 & 0.03 & 0.03 & $0.00172^{<0.021}$ & $0.011^{<0.030}$ & \phs$0.220\!:\pm0.007$ & $3.50\pm0.02$ & Truncated by chip edge\\
560    & J033206.3$-$274537 & $<\!-15.33$ & $-1.00$ & 23.05 & 20.65 & 0.27 & 0.03 & 0.03 & $0^{<0.15}$ & $0.0068^{<0.061}$ & \phs$0.059\pm0.012$ & $3.34\pm0.03$ & \\
563    & J033231.5$-$274624 & $<\!-15.34$ & $-1.00$ & 23.64 & 22.31 & 0.26 & 0.02 & 0.03 & $0.0067^{<0.13}$ & $0^{<0.076}$ & \phs$0.641\pm0.045$ & $3.07\pm0.03$ & \\
593    & J033214.8$-$274403 & $<\!-15.34$ & $-1.00$ & 26.27 & 24.51 & 0.47 & 0.00 & 0.00 & \nodata & \nodata & \phs$0.100\pm0.250$ & $2.66\pm0.08$ & \\
594    & J033209.8$-$274249 & $<\!-15.19$ & $-1.00$ & 22.93 & 20.30 & 0.44 & 0.02 & 0.03 & $0^{<0.0093}$ & $0.025^{+0.0033}_{-0.0032}$ & \phs$0.066\pm0.019$ & $3.61\pm0.03$ & \\
623    & J033228.6$-$274659 & $<\!-15.35$ & $-1.00$ & 26.95 & 26.35 & 0.12 & 0.00 & 0.00 & \nodata & \nodata & \phs$0.156\pm0.236$ & $1.75\pm0.65$ & \\
624    & J033229.3$-$274708 & $<\!-15.33$ & $-1.00$ & 22.68 & 20.36 & 0.33 & 0.03 & 0.03 & $0^{<0.0060}$ & $0^{<0.012}$ & \phs$0.034\pm0.013$ & $3.46\pm0.03$ & \\
626    & J033209.5$-$274758 & $<\!-15.33$ & $-1.00$ & 25.44 & 24.20 & 0.21 & 0.01 & 0.00 & \nodata & \nodata & \phs$0.123\pm0.163$ & $3.20\pm0.09$ & \\
631    & J033215.2$-$274335 & $<\!-15.32$ & $-1.00$ & 25.13 & 22.93 & 0.23 & 0.00 & 0.03 & $0.027^{<0.10}$ & $0.054^{<0.099}$ & \phs$0.038\pm0.065$ & $2.80\pm0.04$ & \\
\enddata
\tablenotetext{a}{Sources detected only in {\sl CXO} soft band (0.5--2~keV) indicated with upper limits}
\tablenotetext{b}{X-ray hardness ratio $\equiv (H-S)/(H+S)$, for hard- and soft-band counts $H$ and $S$}
\tablenotetext{c}{If 0 is not excluded at $>3\sigma$, only $3\sigma$ upper limit is shown; otherwise limits
are $\pm1\sigma$.}
\tablenotetext{d}{{\sl HST}-unresolved sources omitted from 2-D fitting, plotted with $f_{\rm nuc}\equiv f_{\rm tot}$ in Fig.~\ref{ptsrcfig}} 
\end{deluxetable}

\begin{references}

\reference{} Abraham, R.~G., Merrifield, M.~R., Ellis, R.~S., Tanvir, N.~R.,
       \& Brinchmann, J.~1999, MNRAS, 308, 569

\reference{} Barton, E.~J., Geller, M.~J., \& Kenyon, S.~J.~2000, ApJ, 530, 600

\reference{} Bershady, M.~A, \& Jangren, A., \& Conselice, C.~J.~2000, AJ, 119, 2645 (BJC00) 

\reference{} Bertin, E. \& Arnouts, S.~1996, A\&AS, 117, 393

\reference{} Brandt, W.~N., Alexander, D.~M., Hornschemeier, A.~E., Garmire, G.~P.,
       Schneider, D.~P., Barger, A.~J., Bauer, F.~E., Broos, P.~S., Cowie, L.~L.,
       Townsley, L.~K., Burrows, D.~N., Chartas, G., Feigelson, E.~D., Griffiths, R.~E.,
       Nousek, J.~A., \& Sargent, W.~L.~W.~2001, AJ, 122, 2810

\reference{} Casertano, S., de Mello, D., Dickinson, M., Ferguson, H.~C., Fruchter, A.~S., 
       Gonzalez-Lopezlira, R.~A., Heyer, I., Hook, R.~N., Levay, Z., Lucas, R.~A., 
       Mack, J., Makidon, R.~B., Mutchler, M., Smith, T.~E., Stiavelli, M., Wiggs, M.~S., 
       Williams, R.~E.~2000, AJ, 120, 2747

\reference{} Conselice, C.~J., Bershady, M.~A, \& Jangren, A.~2000, ApJ, 529, 886

\reference{} Corbin, M.~R.~2000, ApJ, 536, L73

\reference{} Cowie, L.~L., Barger, A.~J., Bautz, M.~W., Brandt, W.~N., \&
       Garmire, G.~P.~2003, ApJ, 584, L57

\reference{} Dahari, O.~1984, AJ, 89, 966

\reference{} Dickinson, M.~E., Giavalisco, M.~2003, in Proc.~ESO/USM Workshop,
       The Mass of Galaxies at Low and High Redshift, ed.~R.~Bender \&
       A.~Renzini (Berlin: Springer), 324

\reference{} Ferrarese, L. \& Merritt, D.~2000, ApJ, 539, 9

\reference{} Fruchter, A.~S. \& Hook, R.~N.~2001, PASP, 114, 144

\reference{} Gebhardt, K., Kormendy, J., Ho, L.~C., Bender, R., Bower,
        G., Dressler, A., Faber, S.~M., Filippenko, A.~V., Green, R.,
        Grillmair, C., Lauer, T.~R., Magorrian, J., Pinkney, J., Richstone,
        D., \& Tremaine, S.~2000, ApJ, 539, 13

\reference{} Giacconi, R., Zirm, A., Wang, J.~X., Rosati, P., Nonino, M.,
        Tozzi, P., Gilli, R., Mainieri, V., Hasinger, G., Kewley, L.,
        Bergeron, J., Borgani, S., Gilmozzi, R., Grogin, N., Koekemoer, A.,
        Schreier, E., Zheng, W. \& Norman, C.~2002, ApJ, 139, 369

\reference{} Gilli, R., Salvati, M., \& Hasinger, G.~2001, A\&A, 366, 407

\reference{} Graham, A.~W., Erwin, P., Caon, N., Trujillo, I.~2001, ApJ, 563, L11

\reference{} Grogin, N.~A., Conselice, C.~J., Chatzichristou, E., Alexander, 
        D.~M., Bauer, F.~E., Hornschemeier, A.~E., Jogee, S., Koekemoer, A.~M.,
	Laidler, V.~G., Lucas, R.~A., Paolillo, M., Ravindranath, S., 
	Schreier, E.~J., Simmons, B.~D., \& Urry, C.~M.~2003, ApJL, submitted

\reference{} Gunn, J.~E.~1979, in Active Galaxtic Nuclei, ed.~C.~Hazard \& S.~Mitton
        (Cambridge: Cambridge Univ. Press), 213

\reference{} Hasinger, G., Lehmann, I., Giacconi, R., Schmidt, M., Tr\"umper,
        J., \& Zamorani, G.~1999, in Highlights in X-Ray Astronomy,
	ed.~B.~Aschenbach \& M.~J.~Freyberg (MPE Rep.~272; Garching: MPE), 199

\reference{} Hornschemeier, A.~E., Brandt, W.~N., Garmire, G.~P., 
        Schneider, D.~P., Barger, A.~J., Broos, P.~S., Cowie, L.~L., 
        Townsley, L.~K., Bautz, M.~W., Burrows, D.~N., Chartas, G., 
        Feigelson, E.~D., Griffiths, R.~E., Lumb, D., Nousek, J.~A., 
        Ramsey, L.~W., \& Sargent, W.~L.~W.~2001, ApJ, 554, 742

\reference{} Koekemoer, A.~M., Grogin, N.~A., Schreier, E.~J., 
        Giacconi, R., Gilli, R., Kewley, L., Norman, C., Zirm, A., 
        Bergeron, J., Rosati, P., Hasinger, G., Tozzi, P., \& Marconi, A.~2002,
        ApJ, 567, 657 (Paper II)

\reference{} Krist, J.~E. \& Hook, R.~2001, The Tiny Tim User's Guide,
        Version 6.0 (Baltimore: STScI)

\reference{} Lilly, S., Schade, D., Ellis, R., Le Fevre, O., Brinchmann, J., 
        Tresse, L., Abraham, R., Hammer, F., Crampton, D., Colless, M., 
        Glazebrook, K., Mallen-Ornelas, G., \& Broadhurst, T.~1998, ApJ, 500, 75

\reference{} Magorrian, J., Tremaine, S., Richstone, D., Bender, R.,
        Bower, G., Dressler, A., Faber, S.~M., Gebhardt, K., Green, R.,
        Grillmair, C., Kormendy, J., \& Lauer, T.~1998, AJ, 115, 2285

\reference{} Miyaji, T., Hasinger, G., \& Schmidt, M.~2000, A\&A, 353, 25

\reference{} Mushotzky, R., Cowie, L.~L., Barger, A.~J., \& Arnaud, K.~A.~2000,
        Nature, 404, 459

\reference{} Press, W.~H., Flannery, B.~P., Teukolsky, S.~A., \&
        Vetterling, W.~T.~1992, {\it Numerical Recipes in C} (2d ed; New York:
        Cambridge Univ.~Press)

\reference{} Roos, N.~1985, ApJ, 294, 479

\reference{} Rosati, P., Tozzi, R., Giacconi, R., Gilli, R., Hasinger, G.,
        Kewley, L., Mainieri, V., Nonino, M., Norman, C., Szokoly, G.,
        Wang, J.~X., Zirm, A., Bergeron, J., Borgani, S., Gilmozzi, R.,
        Grogin, N., Koekemoer, A., Schreier, E., \& Zheng, W.~2002, ApJ,
        566, 667

\reference{} Schreier, E.~J., Koekemoer, A.~M., Grogin, N.~A., Giacconi, R.,
        Gilli, R., Kewley, L., Norman, C., Hasinger, G., Rosati, P.,
        Marconi, A., Salvati, M., \& Tozzi, P.~2001, ApJ, 560, 127 (Paper I)

\reference{} Schmitt, H.~R.~2001, AJ, 122, 2243

\reference{} Taniguchi, Y.~1999, ApJ, 524, 65

\reference{} Tozzi, P., Rosati, P., Nonino, M., Bergeron, J., Borgani, S.,
        Gilli, R., Gilmozzi, R., Hasinger, G., Grogin, N., Kewley, L.,
        Koekemoer, A., Norman, C., Schreier, E., Shaver, P., Szokoly, G.,
        Wang, J.~X., Zheng, W., Zirm, A. \& Giacconi, R.~2001, ApJ, 562, 42

\reference{} Urry, C.~M. \& Padovani, P.~1995, PASP, 107, 803

\reference{} Walker, I.~R., Mihos, J.~C.,\& Hernquist~L.~1996, ApJ, 460, 121

\reference{} Wilkes, B.~J., Schmidt, G.~D., Cutri, R.~M., Ghosh, H., Hines, D.~C.,
        Nelson, B., \& Smith, P.~S.~2002, ApJ, 564, L65

\end{references}
\end{document}